\pdfoutput=1 
\documentclass{JINST}
\let\ifpdf\relax
\usepackage{tabularx}
\usepackage{acronym}
\usepackage{subfigure}
\usepackage{graphicx}
\usepackage[english]{babel}
\usepackage{subfloat}
\usepackage{cite}
\usepackage{caption}
\usepackage{lineno}

\makeatletter
\let\@fnsymbol\@arabic
\makeatother

\title{Performance of fully instrumented detector planes
of the forward calorimeter of a Linear Collider detector}
 
\author{ 
\begin{center}
{\Large {The FCAL collaboration}}
\end{center}
H.~Abramowicz$^a$, A.~Abusleme$^b$, K.~Afanaciev$^c$, J.~Aguilar$^{d,}$\thanks{Now at Johns Hopkins University, Baltimore, USA}, E.~Alvarez$^b$, D.~Avila$^b$, Y.~Benhammou$^a$, L.~Bortko$^{e,2}$, O.~Borysov{$^a$}, M.~Bergholz$^{e,}$\thanks{Also at Brandenburg University of Technology, Cottbus, Germany}, I.~Bozovic-Jelisavcic$^f$,  E.~Castro$^{e,}$\thanks{Now at DESY, Hamburg, Germany}, G.~Chelkov$^g$,  C.~Coca$^h$, W.~Daniluk$^d$, L.~Dumitru$^h$,  K.~Elsener$^i$, V.~Fadeyev$^j$, M.~Firlej$^k$,  E.~Firu$^l$, T.~Fiutowski$^k$, V.~Ghenescu$^l$, M.~Gostkin$^g$, H.~Henschel$^e$, M.~Idzik{$^k$}, A.~ Ishikawa$^m$, S.~Kananov{$^a$}, S.~Kollowa$^{e,}$\thanks{Now at Institut fuer Kristallzuechtung, Berlin, Germany}, S.~Kotov$^g$, J.~Kotula$^d$, D.~Kozhevnikov$^g$, V.~Kruchonok$^g$,   B.~Krupa{$^d$}, Sz.~Kulis$^{k,}$\thanks{Now at CERN, Geneva, Switzerland},  W.~Lange$^e$, T.~Lesiak{$^d$}, A.~Levy$^a$, I.~Levy$^a$, W.~Lohmann$^{e,2}$, S.~Lukic$^f$, C.~Milke$^j$, J.~Moron$^k$, A.~Moszczynski$^d$,  A.T.~Neagu$^l$, O.~Novgorodova$^{e,}$\thanks{Now at Technical University Dresden, Dresden, Germany},  K.~Oliwa$^d$, M.~Orlandea$^h$,  M.~Pandurovic$^f$, B.~Pawlik$^d$, T.~Preda$^l$, D.~Przyborowski$^k$, O.~Rosenblat{$^a$}, A.~Sailer$^i$, Y.~Sato$^{m,}$\thanks{Now at Nagoya University, Nagoya, Japan}, B.~Schumm$^j$,  S.~Schuwalow$^{e,}$\thanks{Also at University of Hamburg, Hamburg, Germany},  I.~Smiljanic$^f$, P.~Smolyanskiy$^g$,  K.~Swientek$^k$, E.~Teodorescu$^{h,}$\thanks{Now at National Institute for Laser, Plasma and Radiation Physics(INFLPR), Bucharest-Magurele, Romania}, P.~Terlecki$^k$, W.~Wierba$^d$, T.~Wojton{$^d$},  S.~Yamaguchi$^m$, H.~Yamamoto$^m$, L.~Zawiejski$^d$, I.S.~Zgura$^l$, A.~Zhemchugov$^g$ \\
\llap{$^a$}Tel Aviv University, Tel Aviv, Israel\\
\llap{$^b$}Pontificia Universidad Catolica de Chile, Santiago, Chile\\
\llap{$^c$} NC PHEP, Belarusian State University, Minsk, Belarus\\
\llap{$^d$}INP PAN, Cracow, Poland\\
\llap{$^e$}DESY, Zeuthen, Germany\\
\llap{$^f$}Vinca Institute of Nuclear Sciences, University of Belgrade, Serbia\\
\llap{$^g$}JINR, Dubna, Russia\\
\llap{$^h$}IFIN-HH, Bucharest, Romania\\
\llap{$^i$}CERN, Geneva, Switzerland\\
\llap{$^j$}University of California, Santa Cruz, USA\\
\llap{$^k$}AGH University of Science and Technology, Cracow, Poland\\
\llap{$^l$}ISS, Bucharest, Romania\\
\llap{$^m$} Tohoku University, Sendai, Japan\\

 E-mail: \email{levyaron@post.tau.ac.il}
 }

\abstract{
Detector-plane prototypes of the very forward calorimetry of a future detector at an $e^+e^-$ collider have been built and their performance was measured in an electron beam. The detector plane comprises silicon or GaAs pad sensors, dedicated front-end and ADC ASICs, and an FPGA for data concentration. 
Measurements of the signal-to-noise ratio and the response as a function of the position of the sensor are presented.
A deconvolution method is successfully applied, and a comparison of the measured shower shape as a function of the absorber depth with a Monte-Carlo simulation is given.
}


\begin{document}

\section{Introduction}\label{Introduction}

\paragraph{}Future $e^+e^-$ colliders offer a rich experimental program that addresses many open issues in elementary particle physics. For example, the recent discovery of a Higgs boson~\cite{HiggsA,HiggsC} challenges to fully explore the mechanism of Spontaneous Symmetry Breaking. For the moment, two such colliders, distinguished by their acceleration concepts and by their energy reach, are being studied; the International Linear Collider (ILC)~\cite{ILC}, based on superconducting cavities, and the Compact Linear Collider (CLIC)~\cite{CLIC} with the two-beam concept. For the ILC, two types of detectors are under design, the International Large Detector (ILD)~\cite{ILD} and the Silicon Detector (SiD)~\cite{SiD}. Similar concepts are also worked out for CLIC, CLIC\_ILD and CLIC\_SiD~\cite{CLIC,CLIC_ILD}.

 For the luminosity determination, a system of very forward detectors is designed, which can be adapted to any of the detectors at future linear colliders. For that purpose, two special calorimeters~\cite{ilc1} are foreseen in the very forward region of a future  detector, the Luminosity Calorimeter (LumiCal) and the Beam Calorimeter (BeamCal).  The LumiCal will measure the luminosity with a precision of better than 10$^{-3}$ at 500~GeV centre-of-mass  energy and 3$\times 10^{-3}$ at 1~TeV centre-of-mass energy at the ILC, and with a precision of 10$^{-2}$ at CLIC. The BeamCal will perform a bunch-by-bunch estimate of the luminosity and, supplemented by  a pair monitor, assist beam  tuning when included in a fast feedback system~\cite{grah_sapronov}.
\begin{figure}[h!]
  \centering
   \includegraphics[width=0.55\textwidth]{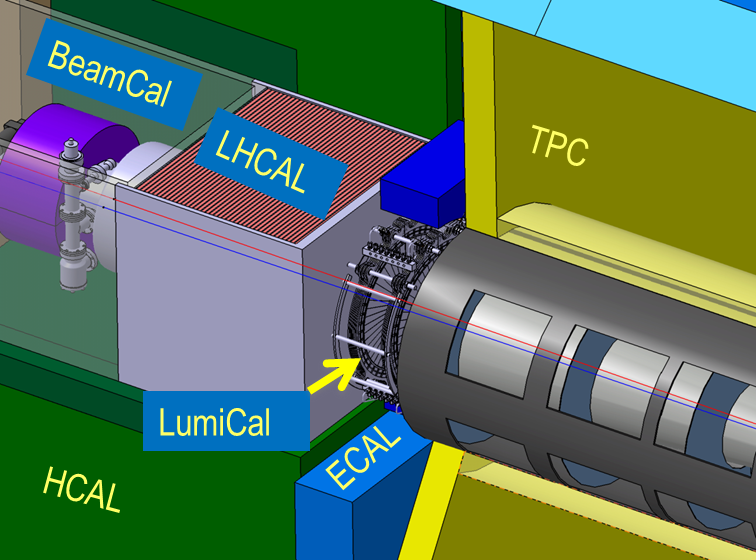}
\caption{The very forward region of the ILD detector. 
LumiCal, BeamCal and LHCAL are carried by 
the support tube for the final focusing quadrupole QD0 and the beam-pipe. 
TPC denotes the central tracking chamber, ECAL the electromagnetic and 
HCAL the hadron calorimeter. }
\label{fig:Forward_structure}
\end{figure}
Both calorimeters extend the detector coverage to low polar angles, 
important e.g. for new particle searches with a missing energy signature~\cite{drugakov}. 
A sketch of the design is shown in Figure~\ref{fig:Forward_structure} for the ILD detector. The LumiCal is positioned in a circular hole of the end-cap electromagnetic calorimeter ECAL.
The BeamCal is placed just in front of the final focus quadrupole.
LumiCal covers polar angles between 31 and 77~mrad and BeamCal, between 5 and 40~mrad.

Both calorimeters consist of 3.5~mm-thick tungsten absorber disks, each corresponding
to around one radiation length, interspersed with sensor layers. Each sensor layer is segmented radially
and azimuthally into pads. The readout rate is driven by the beam-induced background.
Due to the high occupancy originating from beamstrahlung and two-photon processes,  
both calorimeters need a fast readout. Front-end (FE) and Analog-to-Digital Converter (ADC)  Application-Specific Integrated Circuits (ASICs) are placed at the outer radius of the
calorimeters. 
In addition, the lower polar-angle range of BeamCal is exposed to a large flux 
of low energy electrons induced by beamstrahlung, resulting in depositions of up to one 
MGy for a total integrated luminosity of 500 fb$^{-1}$ at 500~GeV~\cite{ilc1}. Hence, radiation hard sensors are needed.

Prototype detector planes assembled with dedicated FE and ADC ASICs for LumiCal and for BeamCal have been built. In this paper, results of their performance following tests in an electron beam are reported. Specifically, the performance to measure minimum ionising particles is investigated. In the collider detectors, it is foreseen to use high energy muons for the alignment of the detector planes, to calibrate the response of all pads regularly, and to account for performance losses, e.g. due to radiation damage. For this purpose, the requirement for the readout is to provide a high enough signal to noise ratio, $S/N$ \textgreater 10, for all pads of BeamCal and LumiCal sensors.

The readout electronics is operated synchronously and asynchronously with the electron beam. The former corresponds to the operation at the ILC with about 300~ns bunch spacing. The latter will be used at CLIC where, due to the bunch spacing of 0.5~ns, all signals will be recorded over a large time window. For both cases a deconvolution method is applied, reducing the amount of data to be recorded. For the application at CLIC, the deconvolution method allows to reconstruct both amplitude and time of a signal. 

\section{Prototype detector planes} 

The detector planes consist of pad sensors instrumented with FE and ADC ASICs. Silicon sensors are used for LumiCal. Since radiation-hard sensors are needed for BeamCal, GaAs and artificial diamond are considered as sensor materials in the design. Both sensors are sufficiently radiation-hard without cooling~\cite{afanaciev2012,cooling}. In the beam tests described in this paper, large area GaAs sensors were used.

\subsection{Silicon sensors}

\begin{figure}[!h]
 \centering
\includegraphics[width=0.85\textwidth]{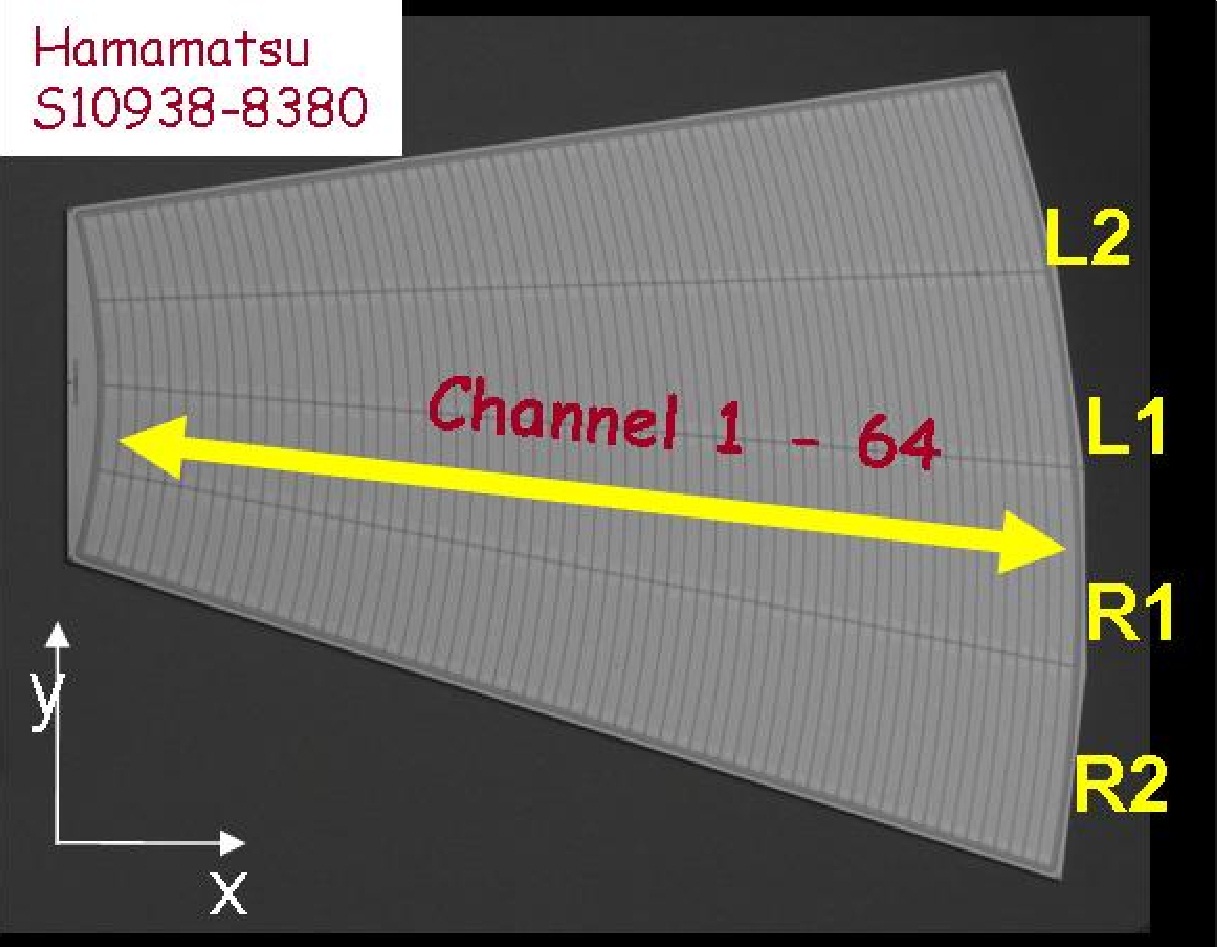}
\caption{A prototype silicon sensor for LumiCal. L1, L2, R1 and R2 are labels for the corresponding sectors. }
\label{figure:Lumical sensor}
\end{figure}

\begin{figure}[!htpb]
    \centering
  \subfigure[]{
    \label{figure:CV_lumi}
      \includegraphics[width=0.45\textwidth]{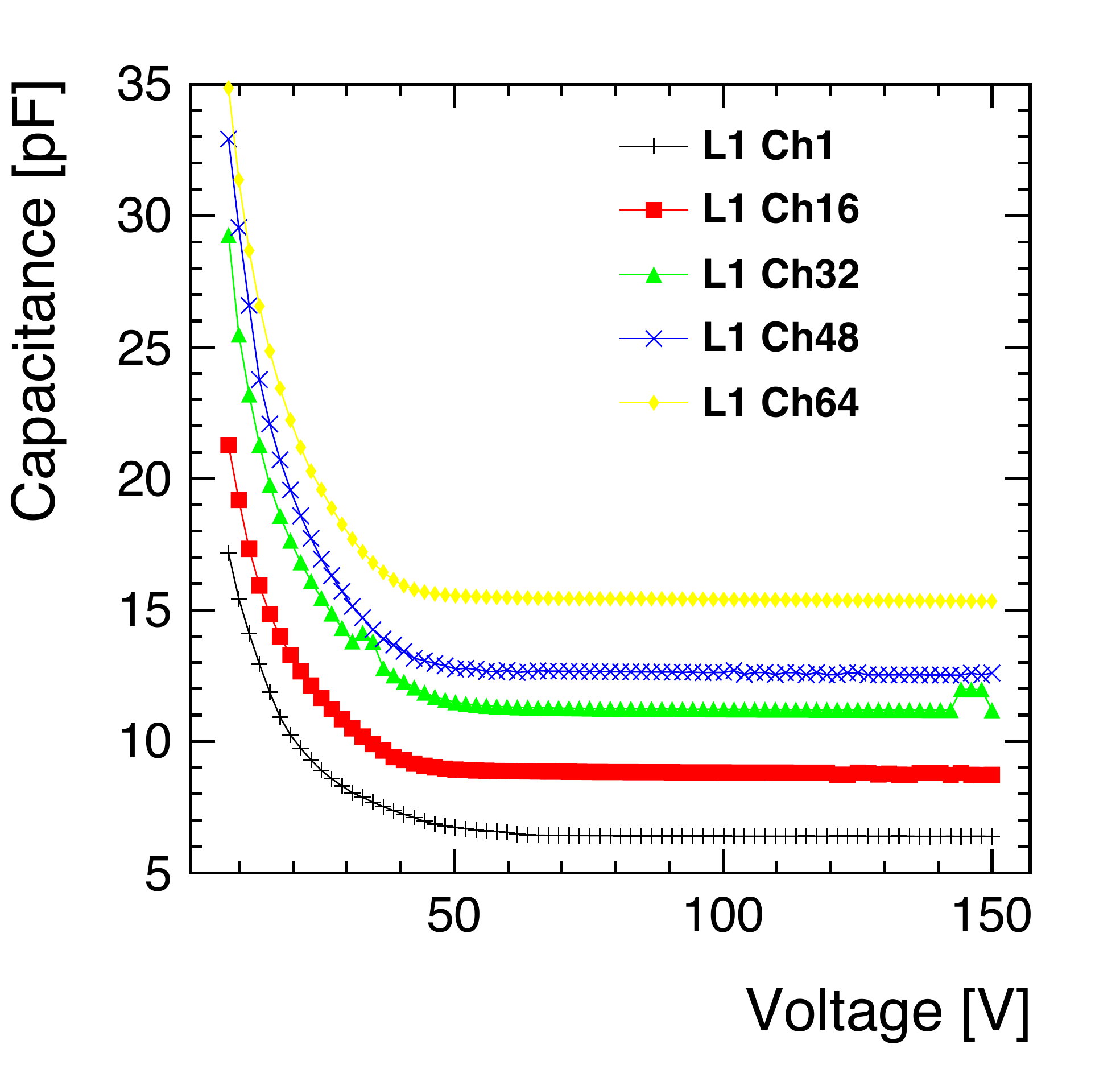}}
   \subfigure[] {
    \label{figure:IV_lumi}
      \includegraphics[width=0.45\textwidth]{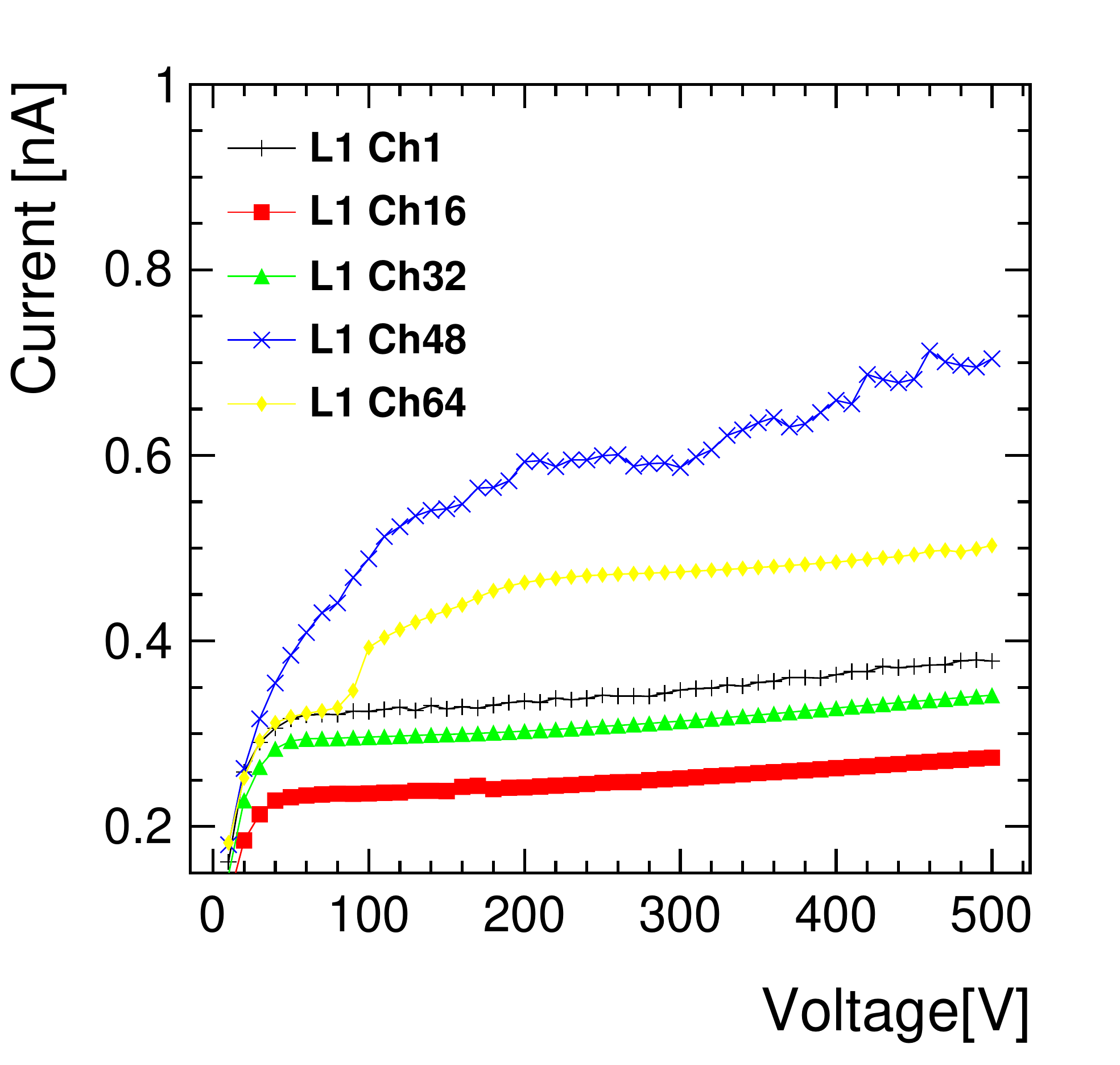}}
      \caption{
    \Subref{figure:CV_lumi} The dependence of the capacitance on the bias voltage for different pads of a silicon sensor.
    \Subref{figure:IV_lumi} The leakage-current dependence on the bias voltage, for the same pads.}
\end{figure}

Prototypes of LumiCal silicon sensors have been designed at the Institute of Nuclear Physics PAN in Cracow~\cite{PAN}
and manufactured by Hamamatsu
Photonics. A picture of a sensor is shown in Figure~\ref{figure:Lumical sensor}.
Its shape is a ring segment of 30$^\circ$ and it contains four sectors, of 7.5$^\circ$ each. The inner radius is 80~mm and the outer radius 195~mm.
The thickness of the n-type silicon bulk is 320~$\mu$m. 
The pitch of the concentric p$^+$ pads is 1.8~mm and
the gap between two pads is 100~$\mu$m. All sensors were characterised for their electrical qualities. 
The capacitance as a function of the bias voltage for different pads is shown in Figure~\ref{figure:CV_lumi}. 
The leakage current of the same pads is shown in Figure~\ref{figure:IV_lumi}. The bias voltage for full depletion ranges between 35 and 60~V
and the leakage currents per pad are below 0.8~nA.   
Pad capacitances between 6~pF (L1~Ch1), 
for the smallest pads, and 16~pF, for the largest pads (L1~Ch64), were measured~\cite{PAN,Itamar}.
The resistivity of the sensor is estimated to be 70~$\Omega$m~\cite{PAN}. 


\subsection{GaAs sensors}

\begin{figure}[!htpb]
 \centering
  \includegraphics[width=0.3\columnwidth, angle=90]{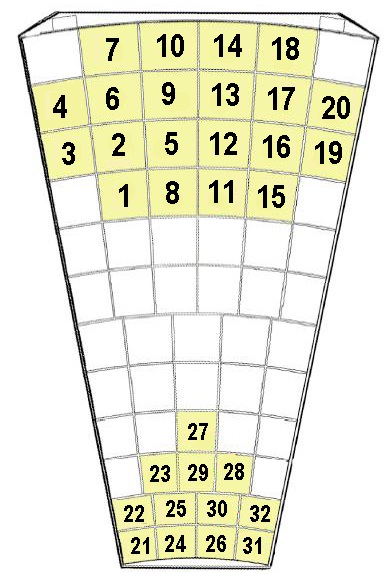}
    \caption{Schematic drawings of a GaAs sensor. Each of the 32 connected pads is identified by a number.}
      \label{figure:GaAs_plane}
\end{figure}

The BeamCal prototype sensors have been produced by Tomsk State University.
The sensors were produced using the liquid encapsulated Czochralski method and were doped with Sn, as a  shallow donor, and compensated with chromium, as deep acceptor. This results in a semi-insulating GaAs material with a resistivity of about 10$^7$~$\Omega$m.
The GaAs sensor has a thickness of 500~$\mu$m. 
Its shape is a ring segment of 22.5$^\circ$. The inner radius is 48~mm and the outer one, 114~mm. 
It is metallised with 1~$\mu$m of Ni on both sides. 
One side has a solid metallisation and the opposite side is segmented into 12~rings and each ring into pads with different area depending 
on the radius, as shown in Figure~\ref{figure:GaAs_plane}.
The total number of pads is 64. The minimal pad area is 18~mm$^2$ and the maximal pad area is 42~mm$^2$.
The gap between two pads is 200~$\mu$m.

\begin{figure}[!h]
    \centering
  \includegraphics[width=0.5\columnwidth]{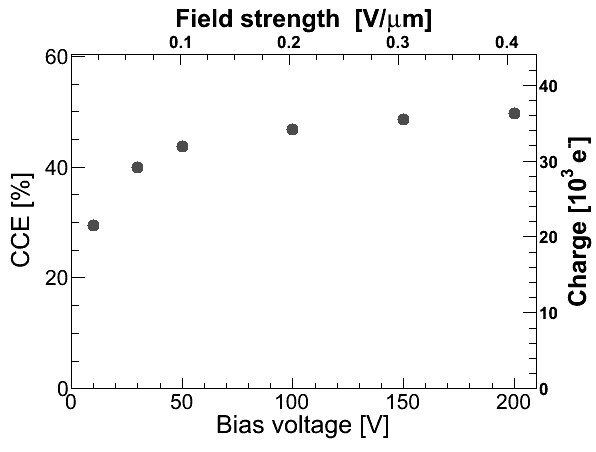} 
    \caption{The CCE and the charge measured as a function of the applied bias voltage for a pad of the GaAs sensor~\cite{OlgaThesis}. The statistical uncertainty is smaller or equal to the size of the dots. From the calibration coefficient a fully correlated systematic uncertainty of 5\% is estimated. }
      \label{figure:gaas_cce_vs_hv_hep}
  \end{figure}

\begin{figure}[h!]
\centering
{\includegraphics[width=0.5\columnwidth]{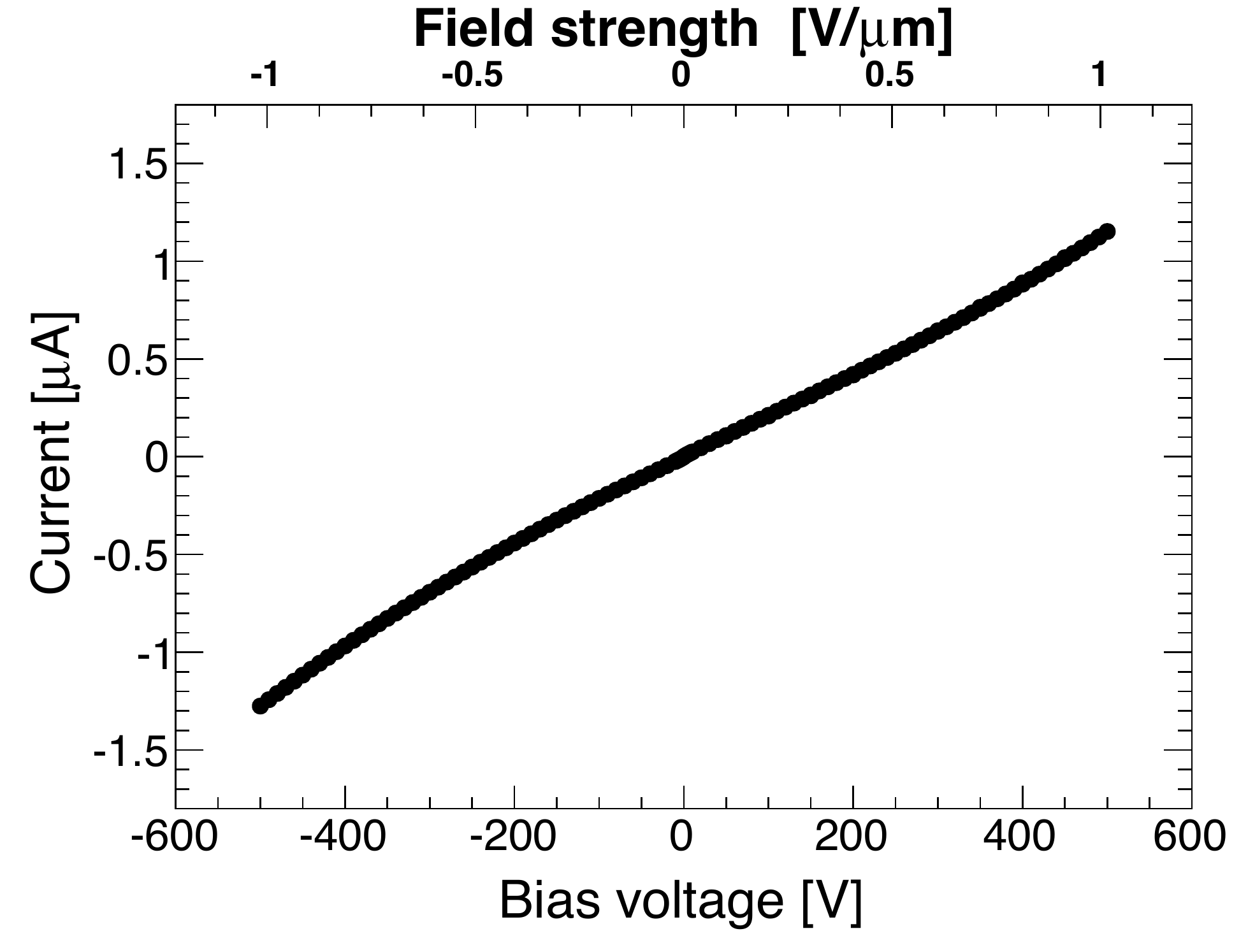} }
     \caption{The leakage current as a function of the applied bias voltage for a pad of the GaAs sensor.}
      \label{figure:gaas_iv}
\end{figure}

Before the beam-test, the sensors were characterised in the laboratory. The charge-collection efficiency  (CCE), defined as the ratio of the measured to the expected signal charge, and the leakage current were measured as a function of the applied voltage and are shown in Figure~\ref{figure:gaas_cce_vs_hv_hep} and Figure~\ref{figure:gaas_iv}, respectively, for a representative pad.
The CCE was measured using signals from electrons of the high-energy tail of a $^{90}$Sr source crossing the sensor and triggered by a scintillator sandwich. Prior to that, the readout chain that was used was calibrated by feeding a known charge into the front-end preamplifier, allowing to measure the signal charge. 
The uncertainty of the calibration factor is estimated to be 5\%.
The expected signal charge is calculated from the average energy-loss of  high energy electrons from  the $^{90}$Sr in the sensor, obtained from a GEANT4 simulation, and the average energy of electron-hole pair creation. It amounts to about 12~fC.
 The CCE saturates at 50\% at a field 
strength of 0.2~V/$\mu$m for all pads. This is expected due to the properties of the compensated material, the hole lifetime is very low and the signal is predominantly generated by electron drift. As a consequence, the maximum reachable CCE of the GaAs sensor is 50\% for particles fully crossing the sensor.
 Pad capacitances vary between 5~pF for the smallest pads and 
12~pF for the largest pads.

Previously, a study was done to determine the radiation hardness of the compensated GaAs sensors in a 10~MeV electron beam~\cite{afanaciev2012}. The charge-collection efficiency was measured during irradiation. It drops by a factor of 10, but a signal from a minimum ionising particle 
is still visible after an absorbed dose of 1.5~MGy. The leakage current of a pad  before 
irradiation was about 0.4~$\mu$A at an electric-field strength of 0.4~V/$\mu$m. After exposure to a dose of 1.5~MGy, the leakage current increased\footnote{The sensors of BeamCal need to be replaced after a few years of running ILC at 500~GeV at design luminosity.} to about 1~$\mu$A. 

\subsection{Sensor planes}

Thin printed circuit boards (PCBs) with copper traces were used as fan-outs.
They were glued to the sensors.
The fan-out itself was screwed to an additional PCB carriage with a window
corresponding to the sensor shape. 
Fan-out traces at one end were bonded to the connector of the readout electronics board and at the other end to the pads through small holes.
High voltage was applied to the fully metallised side of
the sensor.
Finally, the fully assembled prototypes were put into a light-tight shielding
box.

\section{Readout electronics}

The readout electronics were developed according to the specifications 
imposed by operating conditions and constraints~\cite{VFCAL_memo}. 
The FE ASICs were designed to work in two modes: the physics mode and the calibration mode.
In the physics mode (low gain) the detector should be sensitive to electromagnetic showers
resulting in high energy deposition and the FE ASICs should process signals up
 to almost 10~pC per channel.  
In the calibration mode (high gain) it should detect signals from relativistic muons, hereafter referred to as
minimum ionizing particles (MIPs) to be used for calibration and alignment. The expected signal charges are 4~fC for the silicon sensor and 6~fC for the GaAs sensor.
The proposed sensor geometry results in a  capacitive load
  (sensor and fanout) between 5~pF~-~35~pF connected to a single FE channel. 
Because of high expected occupancy\footnote{The average occupancy for BeamCal is above 50\% and for LumiCal a few \%, but it varies locally.} per channel, the FE ASIC 
should be fast enough to process signals from subsequent beam bunches which, for the ILC, are separated in time by about 300~ns.
The simulations of LumiCal and BeamCal indicate that the shower reconstruction needs  a 10-bit ADC.
Severe requirements on the power dissipation of readout electronics 
may be strongly relaxed if the power is switched off between bunch trains. 
This is feasible for the ILC,  since a 
200~ms pause is foreseen after each 1~ms long bunch train~\cite{beam}.

\begin{figure}[htb!]
  \centerline{\includegraphics[width=1\columnwidth]{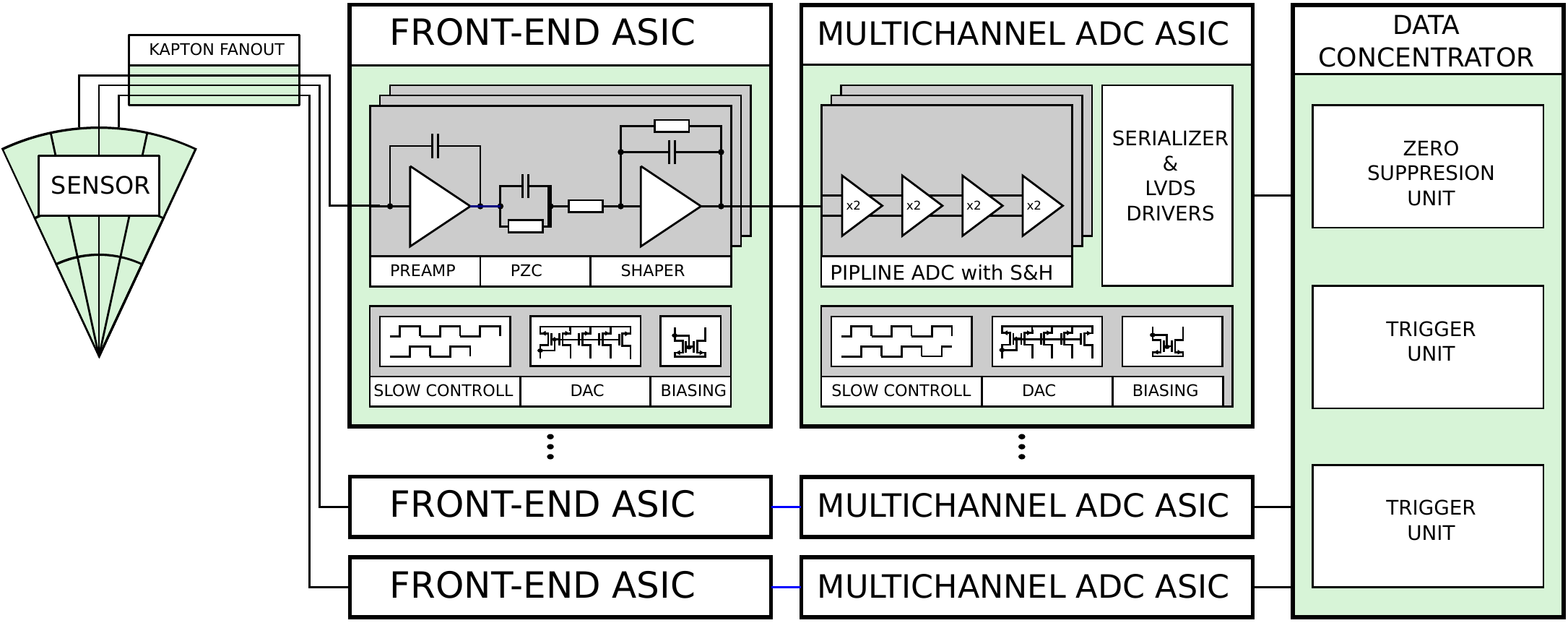}}
\caption{Block diagram of the readout-chain.\label{fig:fcal_lumical_system}}
\end{figure}

From the above specifications, a general concept of the readout chain
was developed, as shown in Figure~\ref{fig:fcal_lumical_system}. The main blocks of the signal-processing chain are sensor, FE ASIC, ADC ASIC and a Field Programmable Gate Array (FPGA) based data concentrator.

\subsection{Front-end electronics }
\label{fcal_lumical_fe}

\begin{figure}[h!]
  \begin{minipage}[t]{0.5\linewidth}
    \centerline{\includegraphics[width=1.35\columnwidth]{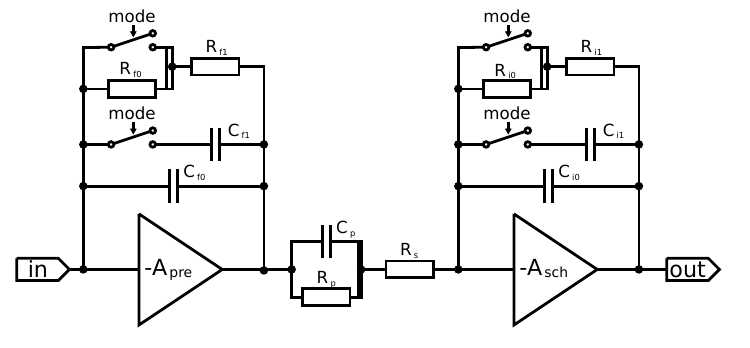}}
    \caption{Block diagram of a front-end channel.}
    \label{fig:frontend}
  \end{minipage}
  \hspace*{0.02\linewidth}
  \begin{minipage}[t]{0.5\linewidth}
    \centering \includegraphics[width=0.8\columnwidth]{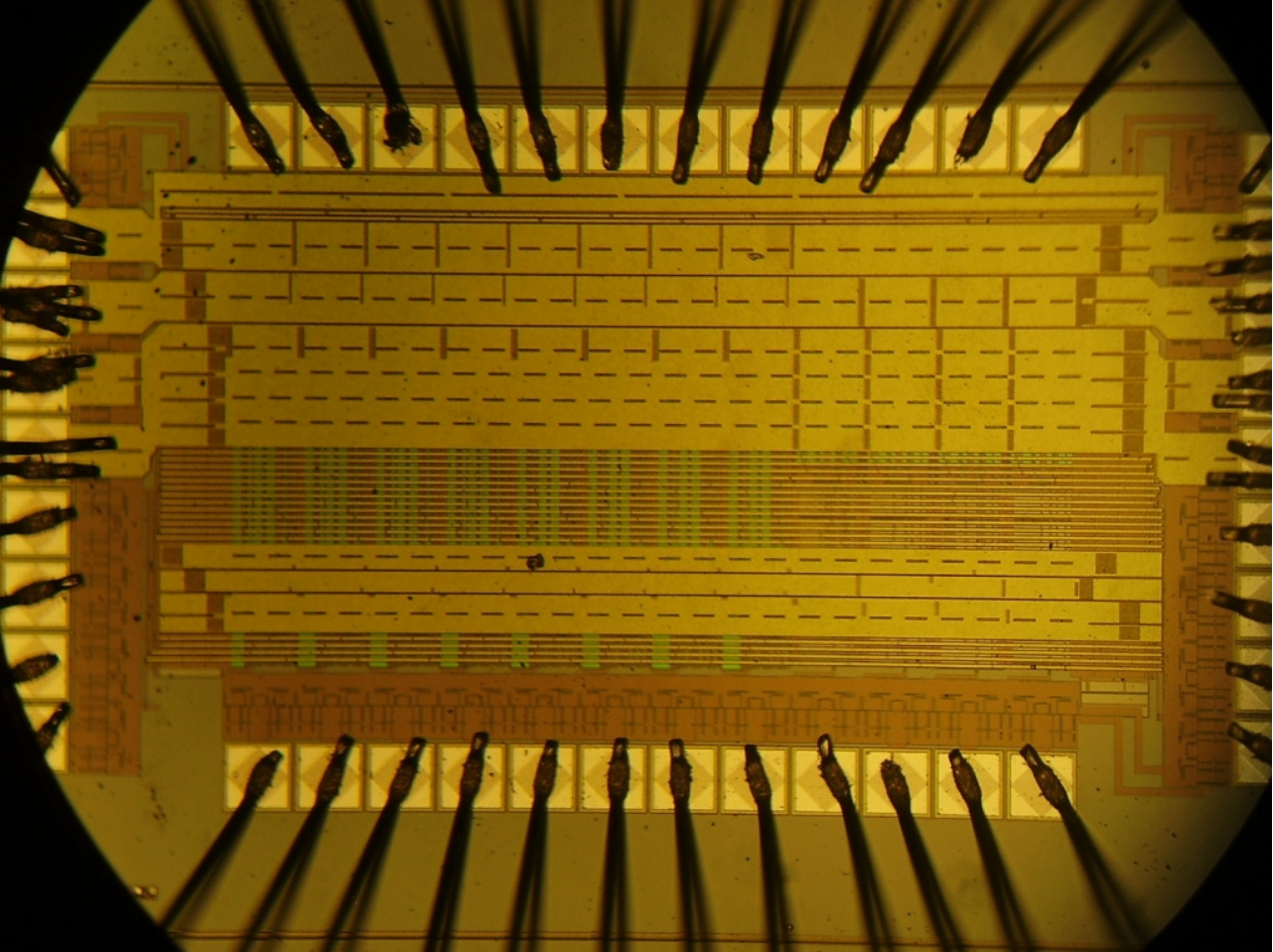}
     \caption{Micrograph of a front-end ASIC.}\label{fig:frontend_photo}
  \end{minipage}
\end{figure}

To fulfill the above requirements, the FE ASICs architecture  comprising a charge sensitive amplifier, 
a pole-zero cancellation circuit (PZC) and a shaper was chosen~\cite{lumi_frontend}, as shown in Figure~\ref{fig:frontend}.
In order to cope with large input charges in the physics mode and the small ones 
in the calibration mode, a variable gain in the charge amplifier and 
shaper was implemented. The ``mode'' switch in Figure~\ref{fig:frontend} changes effective values of the feedback circuit components $R_f$, $C_f$, $R_i$, $C_i$ 
and so changes the transimpedance gain of the FE ASIC.
In the preamplifier feedback and the PZC, two versions of resistance were implemented; a standard passive resistance, henceforth referred to as $R_f$ feedback, and a MOS transistor working in a linear region, henceforth referred to as $MOS$ feedback. In addition, for the calibration mode, the preamplifier gain in the $MOS$ feedback version was set two times higher than in the $R_f$ feedback version. By setting the PZC parameters properly  ($C_f R_f = C_p R_p$) and equalising 
shaping time constants ($C_i R_i = C_p (R_p || R_s)$) , one obtains the first-order shaping, equivalent to a CR-RC filter, with a peaking time $T_{peak}=C_iR_i$, which was set close to 60~ns.

The prototype ASIC, containing eight front-end channels (four with $R_f$ feedback, four with $MOS$ feedback), was designed and fabricated in 0.35~$\mu$m four-metal two-poly CMOS technology. 
The area occupied by a single channel is $630~\mu$m $\times$ 100~$\mu$m. 
The micrograph of the prototype glued and bonded on the PCB is shown in Figure \ref{fig:frontend_photo}. 
Systematic measurements of the essential parameters like gain, noise, high count-rate performance and crosstalk,
were performed as described in detail in Ref.~\cite{lumi_frontend}, confirming the expectations. 

\subsection{Multichannel ADC }

To apply an analog-to-digital conversion in each FE channel, a dedicated low power, small area, multichannel ADC is needed.
For an ILC detector, a sampling rate of about 3~MS/s will be sufficient while for the beam-test purpose a much faster ADC, allowing few samples per pulse,  is beneficial. To meet both requirements, a general purpose 
variable-sampling-rate ADC with scalable power consumption was developed. 

\begin{figure}[h!]
  \begin{minipage}[t]{0.5\linewidth}
    \centering
    \includegraphics[width=1\columnwidth]{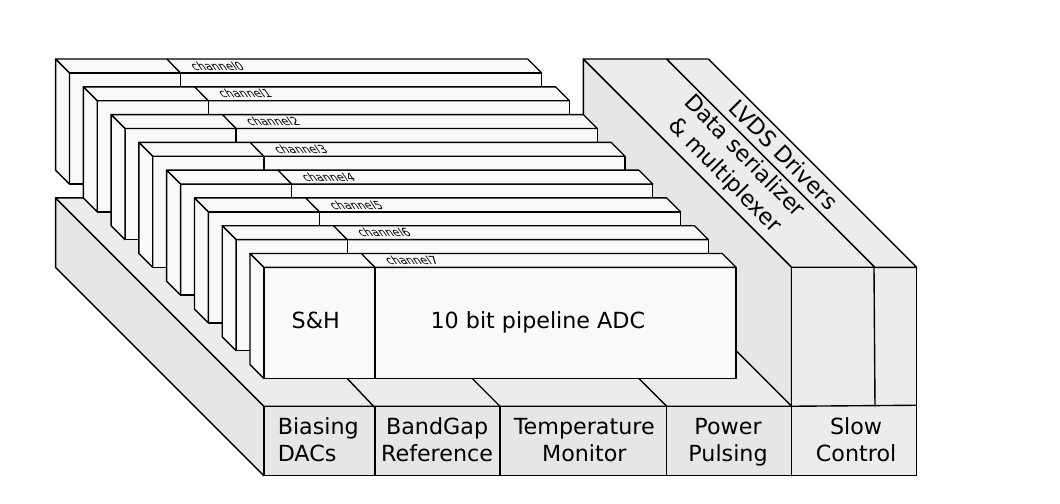}
    \caption{A multichannel ADC block diagram.}
    \label{fig:multi_adc_diagram}
  \end{minipage}
  \hspace*{0.01\linewidth}
  \begin{minipage}[t]{0.5\linewidth}
    \centering
    \includegraphics[width=0.5\columnwidth,angle=90]{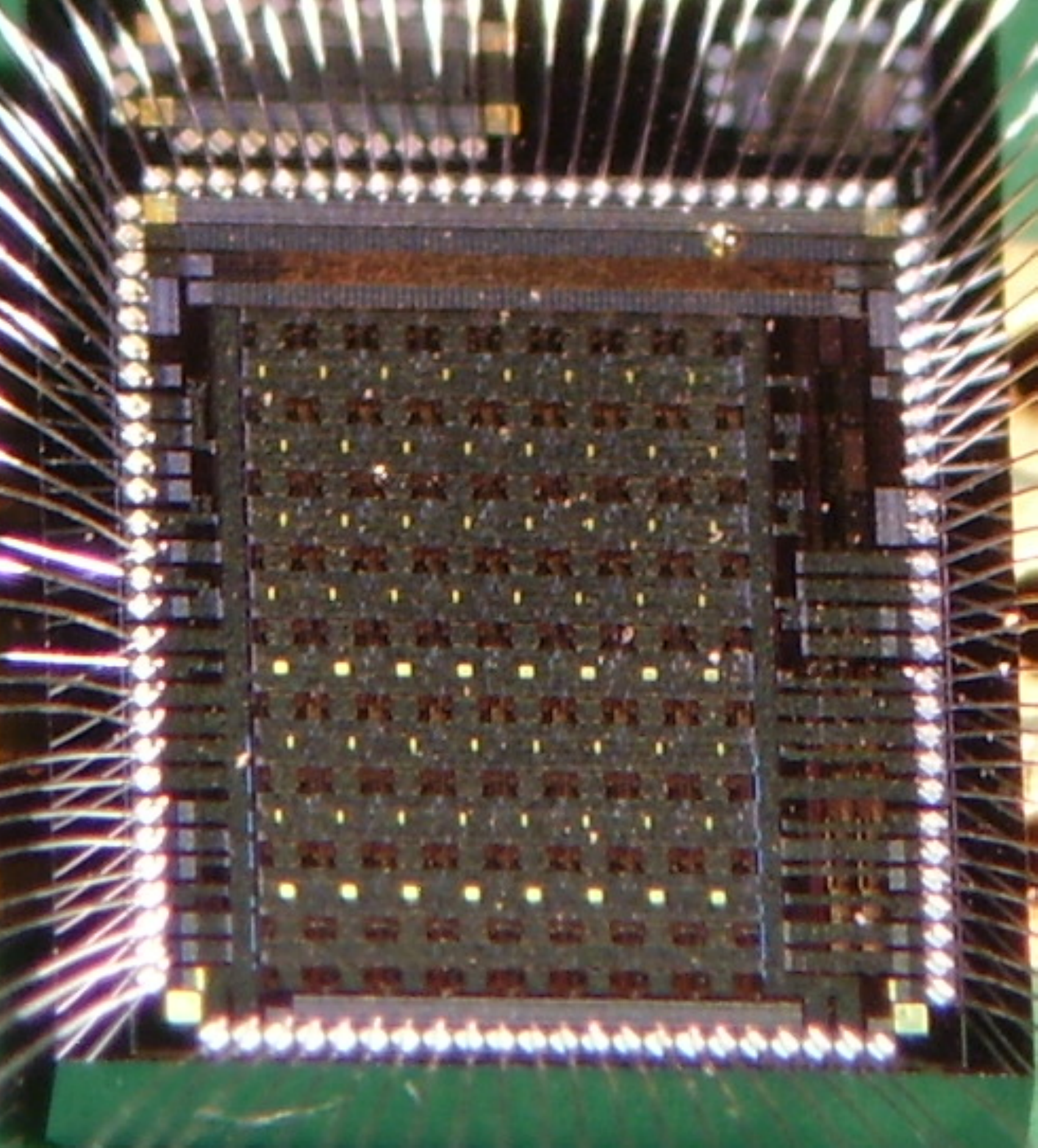}
    \caption{A micrograph of the ADC ASIC.}
    \label{fig:adc_photo}
  \end{minipage}
\end{figure}

The block diagram of the designed multichannel ADC ASIC~\cite{lumi_adc} is shown in Figure~\ref{fig:multi_adc_diagram}. 
It comprises eight 10-bit ADCs with variable power and sampling frequency up to 24 MS/s, a configurable digital serialiser circuit, 
fast variable power Low Voltage Differential Signaling (LVDS) I/O circuits, a set of Digital-to-Analog converters for automatic internal current and voltage control, 
a precise bandgap voltage reference and a temperature sensor. The only external signals needed, apart from a power supply, are reference voltages to set the range of the ADC input signal.
The developed digitiser comprises also the power-pulsing functionality. About 10 ADC clock periods are needed to restart the correct ADC operation after the pause.  

A micrograph of the multichannel ADC prototype, glued and bonded on a PCB, is shown in Figure~\ref{fig:adc_photo}. 
The active size of the ASIC is $3.17~{\rm mm} \times 2.59~{\rm mm}$. Eight ADC channels are placed 
in parallel with 200~$\mu$m pitch and are followed by the serialiser and LVDS pads, while 
the analog and digital peripheral circuits are on the ASIC sides.  
The measurements of performance, comprising static and dynamic parameters, power consumption, and crosstalk, were carried out as described in detail in Ref.~\cite{lumi_adc}. An effective  resolution (ENOB) of 9.7 bits was measured in the whole sampling frequency range and for the input frequency up to the Nyquist frequency.

\subsection{Detector plane}

\begin{figure}[h!]

\centering
\includegraphics[width=0.35\columnwidth,angle=90]{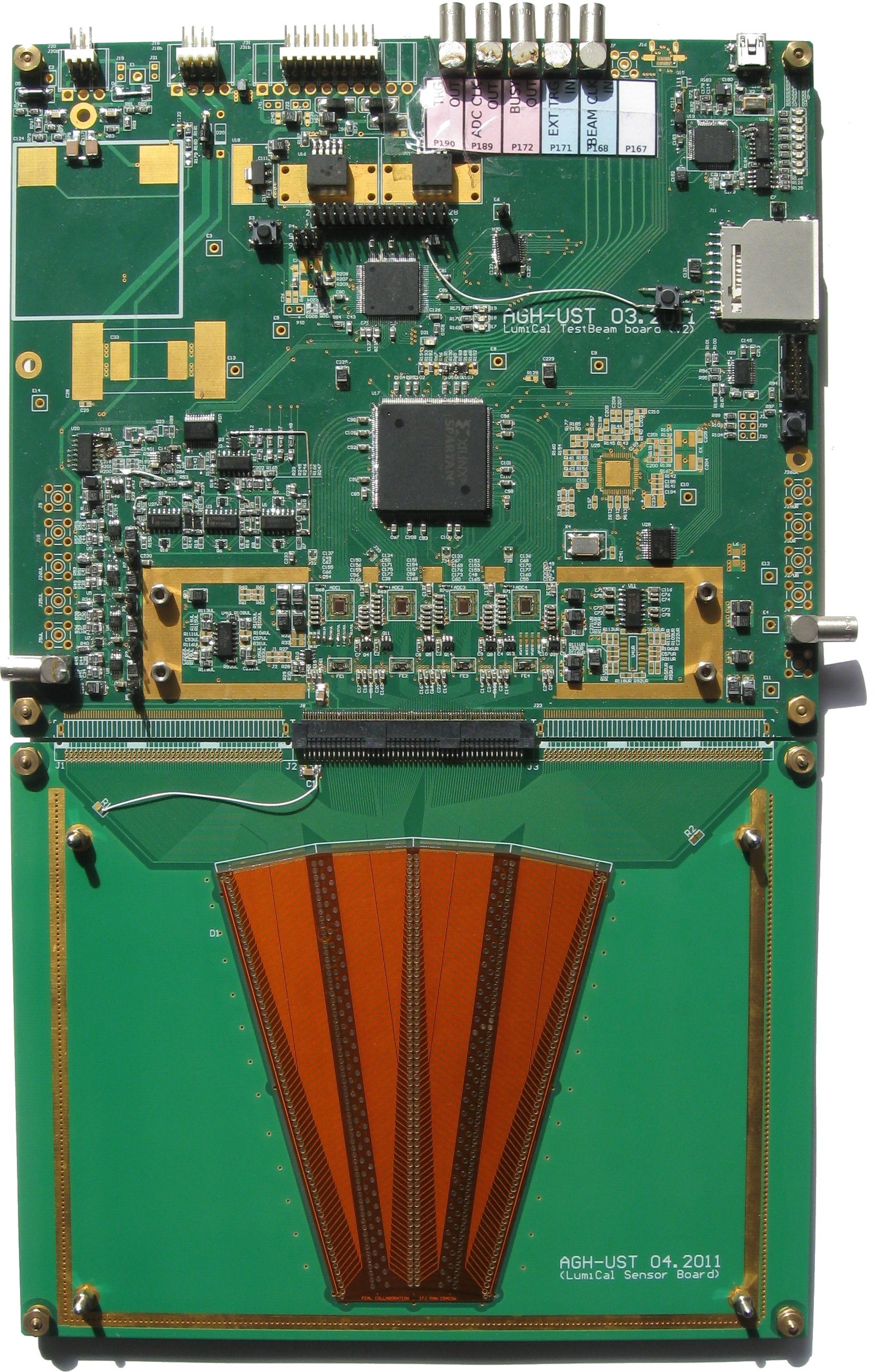}
\caption{Photograph of a detector plane with a silicon sensor connected. The total size amounts to 
34$\times$21~cm$^2$.}
\label{fig:fcal_lumical_module_photo}
\end{figure}
A fully assembled detector plane is shown in Figure~\ref{fig:fcal_lumical_module_photo}~\cite{fcal_readout_board}. The prototype system was fabricated and assembled on a 6-layer PCB. 
In order to increase the system flexibility and to allow it to  operate with different sensors, the signal is sent to the FE ASICs through a multi-way connector.
The signal is then amplified, shaped and continuously digitised.
Four pairs of FE and ADC ASICs, eight~channels each ( 32 channels in total)  are placed on the readout board.  
Power pulsing for the ADC was done by sending a control signal, while for the FE ASIC, additional external switches disconnecting the bias currents were used. 
To reconstruct the signal amplitude and time, and to perform pile-up studies, special attention was given to ensure a sufficiently high ADC sampling-rate and very high internal-data throughput between the ADC and the FPGA. The signal is sampled with a 20~MS/s rate, and digitised with 10-bit resolution, resulting in a raw data stream of about 6.4~Gb/s. To fulfil the high throughput requirements, the digital back-end was implemented on a low-cost, 
high-density Spartan XC3500E FPGA~\cite{xilinx} and a ATxmega128A1 microcontroller~\cite{atmel}.
The digitised data stream is continuously recorded in a buffer inside the FPGA.
When a trigger is received, the acquisition is interrupted and the microcontroller firmware builds an event packet and transmits it to a host PC.

To verify and to quantify the multichannel readout performance, various laboratory measurements of different system sections  and the complete readout chain were carried out.
For the analog part, the measurements of input dynamic range, gain, and noise and their uniformities were also performed. 
For the digital part, the data transmission rate and the event trigger rate in different readout conditions were measured. 
The performance of the complete readout system in a self-triggering mode was verified with cosmic rays. Additionally, measurements of power consumption and thermal system behaviour in the power-pulsing mode were done. These measurements are described in Ref.~\cite{fcal_readout_board}.  
After successful completion of the laboratory tests, the developed readout system was installed in the beam-tests.

\section{Beam-test instrumentation and analysis tools}

During 2010 and 2011, the FCAL collaboration performed three beam-tests.
These were the first tests of the LumiCal silicon- and the BeamCal GaAs-sensors prototypes equipped with a  readout chain.  In 2010, the readout comprised a board with two front-end ASICs (16 channels) and an external ADC. In 2011, the complete 32 channel readout module described in Section 3 was used. 
The detector module performance results include basic signal studies, signal-to-noise measurements, signal deconvolution, response uniformity over the sensor area and shower development measurements with tungsten absorber plates upstream of the sensors.

 \subsection{Beam-tests setup}

\begin{figure}[h]
    \centering
      \includegraphics[width=1.1\textwidth]{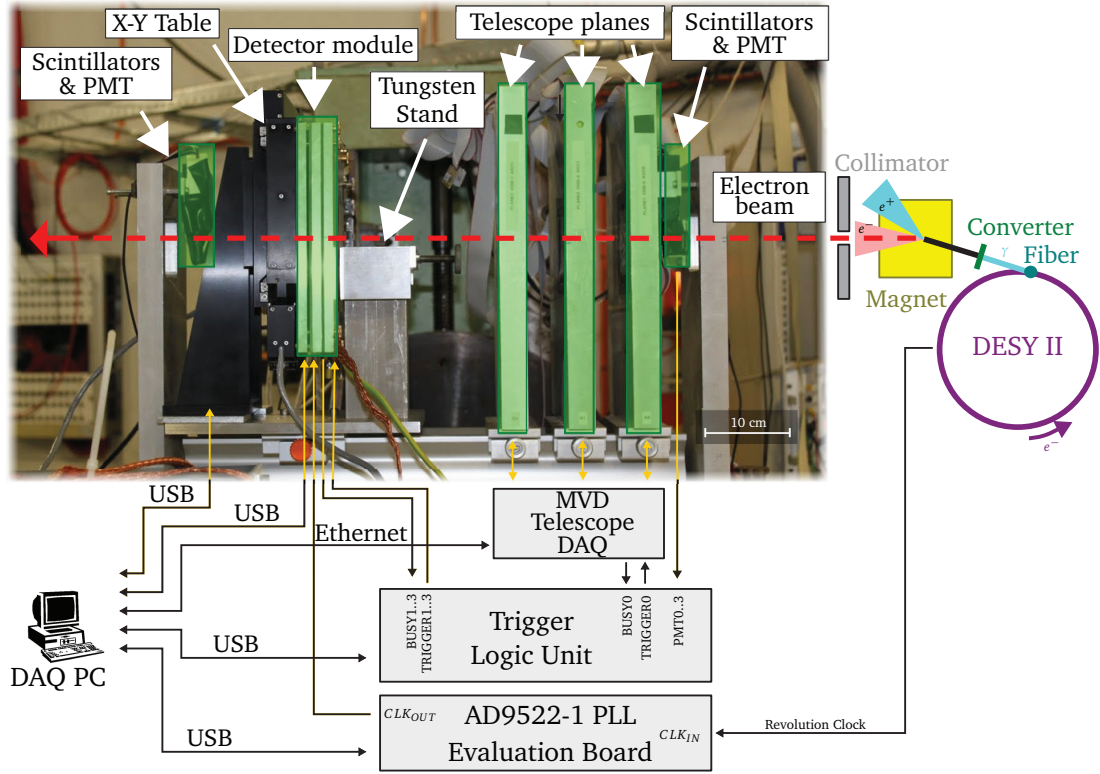}
      \caption{ Scheme of the beam-test setup from the 2011 test period. The distance between the telescope planes is 80~mm, and the distance between the last telescope plane and the detector module is 230~mm.}
    \label{figure:tb2011_setup_sch}
\end{figure}
The beam tests were performed at the DESY-II Synchrotron  with secondary electrons of energies between 2 and 4.5~GeV  and a few hundred particles per second, adapted to the readout speed of the Data Acquisition System (DAQ).  The scheme of the setup for the 2011 test period is shown in Figure~\ref{figure:tb2011_setup_sch}.

The particles crossed three planes of  the ZEUS MVD telescope~\cite{mvd1}.
The horizontal and vertical beam size was about 5x5 ${\rm mm}^2$.
Scintillation counters upstream and downstream of the setup were used in coincidence as trigger counters. 
In the 2010 tests, the detector plane was mounted between the second and the third MVD plane inside a shielded PCB box.
In 2011, it was installed downstream of the three MVD planes, as shown in Figure~\ref{figure:tb2011_setup_sch}.
The detector plane was mounted on a remotely movable X-Y table. In front of the detector plane, a tungsten block of different thickness was placed in several data-taking runs.

\subsection{Telescope}

Each plane is made of two 300 $\mu$m thick single-sided silicon strip sensors of 32x32 mm$^2$ 
size with a strip pitch of 25 $\mu$m and a readout pitch of 50 $\mu$m. 
The strip directions of the two sensors in a module are perpendicular to each other. 
Using the data from each telescope plane, the track of the beam electron can be 
determined and the point where it hits the device under test (DUT) can be accurately calculated. 

A track reconstruction algorithm was developed using the hit-point coordinates of each telescope plane. 
The centroids of the clusters of electric charge produced by the 
passage of incident electrons in the telescope planes were stored.
The algorithm used the following condition:
\begin{equation}
 min \left(\sum_{i=1}^{3}((x_{ip} - x_{im})^{2}+(y_{ip} - y_{im})^{2})\right), 
\label{equ:minimum}
\end{equation}
where $(x_{im}, y_{im})  $ are the measured coordinates and $ (x_{ip}, y_{ip})$ are the predicted coordinates
given by the intersection point of a line with each telescope plane.

For the following analysis, only triggers with one reconstructed track and no additional hits in the MVD planes are used. The fraction of such triggers is about 30\%.

\begin{figure}[ht!]
    \centering
  \subfigure[]{
    \label{fig:resX2_ref_1.pdf}
      \includegraphics[width=0.48\textwidth]{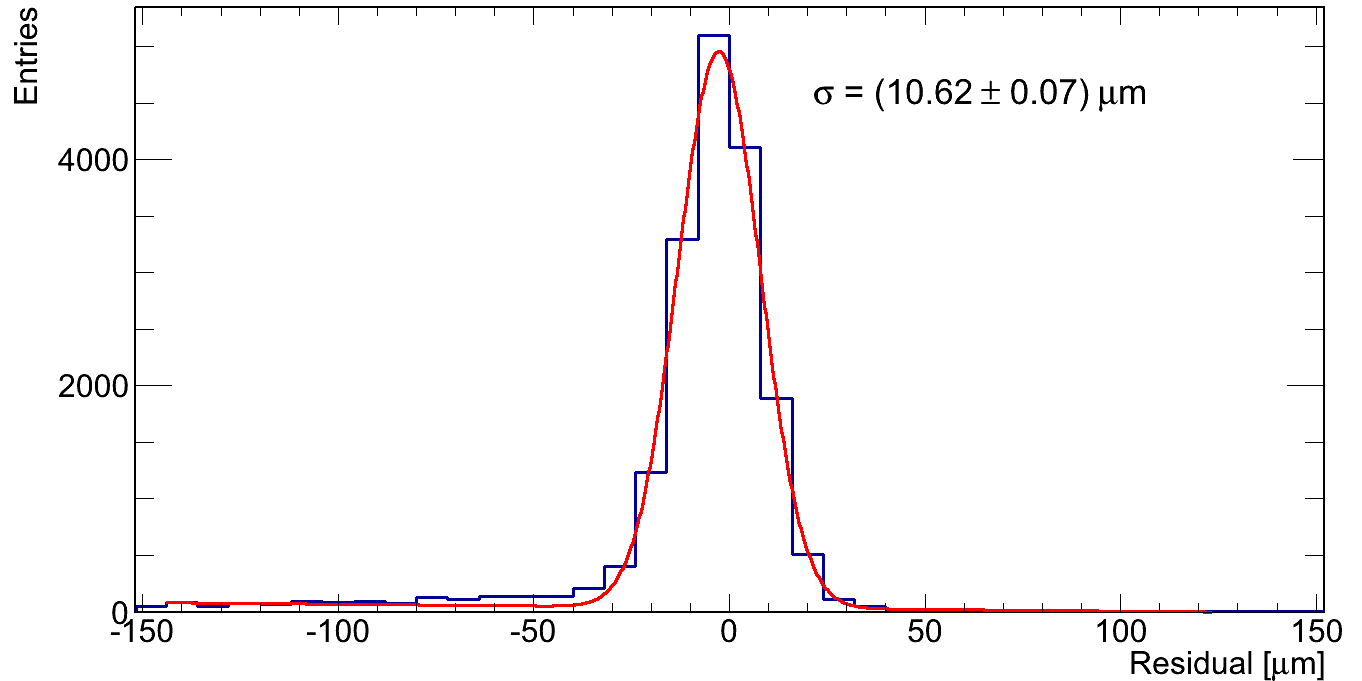}}
   \subfigure[] {
    \label{fig:resY1_ref_1.pdf}
      \includegraphics[width=0.485\textwidth]{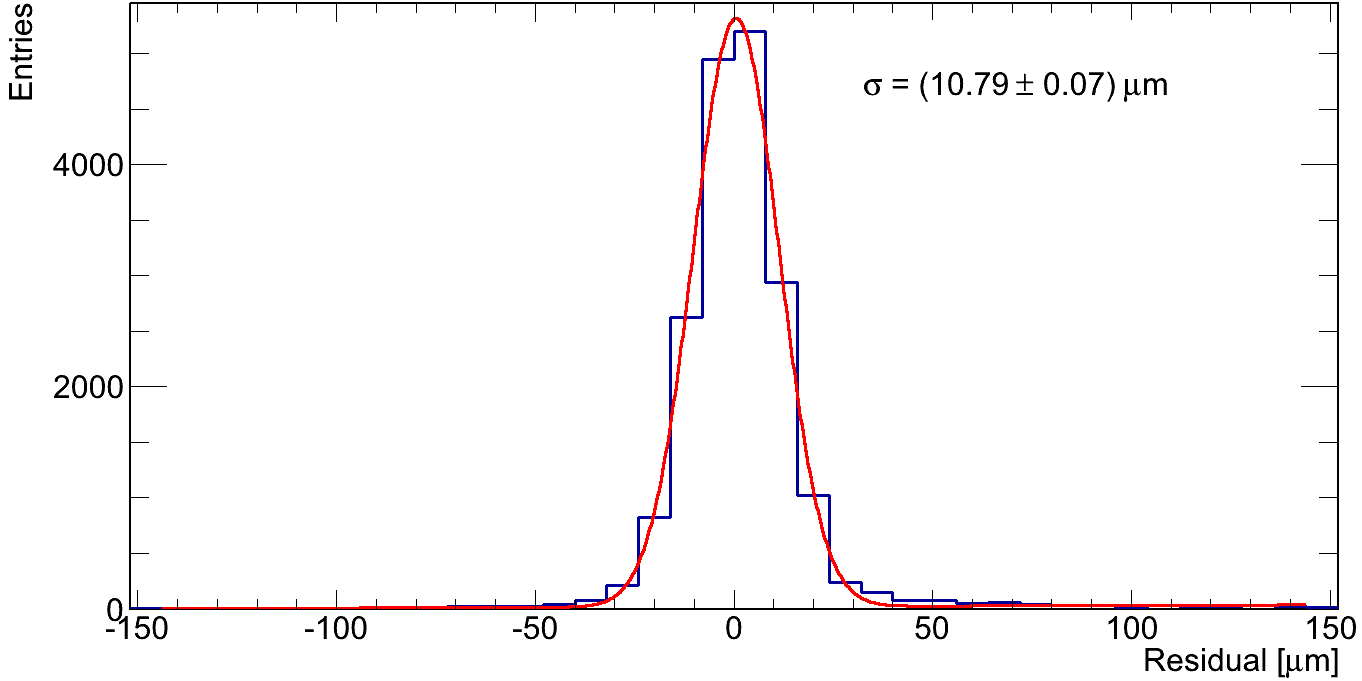}}
      \caption{ Distribution of residuals in 
    \Subref{fig:resX2_ref_1.pdf}  the X direction,
    \Subref{fig:resY1_ref_1.pdf} the Y direction.
In the legend, the quoted $\sigma$ values are the results of fitting a Gaussian to the residual distribution (shown as a continuous line).}
\end{figure}

Special runs were used to align the MVD planes. The width of the residuals of the intersection point on each telescope 
plane was determined to be about 11 $\mu$m, as shown
in Figures~\ref{fig:resX2_ref_1.pdf} and~\ref{fig:resY1_ref_1.pdf}.

After a survey, it was still necessary to make some relative
shifts between telescope planes, however the shifts were less than 100~$\mu$m.

\subsection{Data Acquisition System (DAQ)}

During the 2010 test,
the DAQ was composed of two separate systems,   
that of the MVD and that of the detector planes.
Each system recorded and stored the data independently.
A veto scheme was used to ensure that both DAQ systems
acquire the same number of events and the event building was performed off-line.

For the 2011 tests the EUDAQ~\cite{EUDAQ}, a global DAQ software, was used to collect the data from the detectors and to stream simultaneously  telescope data to the same file.
A trigger logic unit (TLU)~\cite{TLU} was used to generate and distribute a trigger to sub-detectors.
Beside acting as a coincidence unit,  the TLU provided for each trigger an identification number. 
This number was used  during the data analysis to synchronise  between events in a robust way.

To enable the detector-module testing in both synchronous and asynchronous operation modes,
the beam revolution clock was delivered from the accelerator to the test setup to synchronise it with the ADC clock.

\subsection{Deconvolution}
\label{chp:deconvolution}

Data were taken with the ADC ASIC sampling the pulse with a frequency of 20~MHz, and 32 samples per channel were recorded in the trigger time window. In the synchronous mode, the ADC clock was synchronised with the accelerator clock, ensuring that the detector signal evolves always at the same time within the recorded window. In the asynchronous mode, the ADC clock is free running. For both operation modes, the deconvolution is applied as described below~\cite{bienz,kulis_dec}.



For a unit pulse input in a sensor, a semi-Gaussian  response $V_{sh}(s)= \frac{1}{(s+ 1 / \tau )^2} $,
with time constant~$\tau$, is obtained at the output of a CR-RC shaper.
To reconstruct the original sensor signal, a deconvolution filter with a transfer function,
\begin{equation}
D(s) = \frac{1}{V_{sh}(s)} = { (s+ 1 / \tau )^2} ,
\label{dec_form_s}
\end{equation}
is applied.
The discrete time  implementation of such a filter in a digital domain  
may be obtained using the $Z$ transform.
Using the pole-zero mapping, each pole or 
zero in the S plane is replaced by its mapped $z$ position according to $z=e^{sT_{smp}}$, 
where $T_{smp}$ is the sampling period. The formula (\ref{dec_form_s}) transforms to
\begin{equation}
\vspace{-0.1 cm}
D(z) = 1 - 2e^{-{T_{smp}}/{\tau}}\ {z^{-1}} + e ^ {-2 {T_{smp}} / {\tau}}\ {z^{-2}} ,
\label{dec_form_lap}
\end{equation}
where $z^{-1}$ is a unit delay. 
From Eq. (\ref{dec_form_lap}), the expression for deconvolution filtering in the time domain is obtained as
\begin{equation}
d(t_i) = Z^{-1} \left ( D(z) \right ) = V_{sh}(t_i)  - 2e^{-{T_{smp}} / {\tau}} V_{sh}(t_{i-1}) + e ^ {-2 {T_{smp}} / {\tau}} V_{sh}(t_{i-2}).
\label{dec_form}
\end{equation}
It may be noticed that the deconvolution filter is very light, 
requiring only two multiplications and three additions. 

\begin{figure}[h]
\begin{center}
\includegraphics{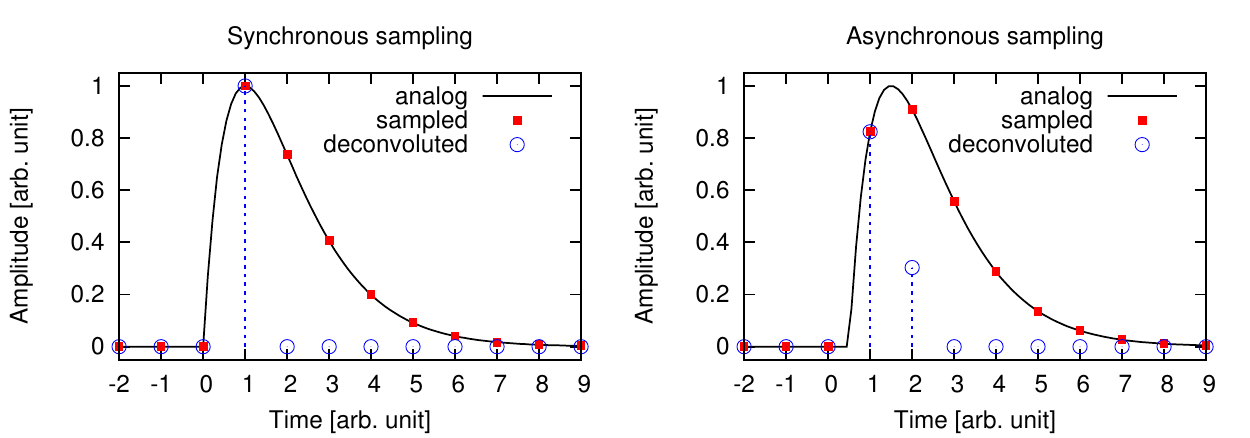}
\caption{\label{dec_resp_fig} Deconvolution filter response ($ T_{smp}=T_{peak} = 1, amp =1  $) for a 
pulse synchronous  (left) and  asynchronous (right) with the ADC clock.}
\end{center}
\end{figure}

In Figure~\ref{dec_resp_fig}, an example of the deconvolution filter response to a sampled shaper output is shown. 
The deconvolution produces only one non-zero sample (Figure~\ref{dec_resp_fig} (left)) with an amplitude proportional to the input pulse. 
This is the case only when the input pulse is synchronised with the ADC clock. 
In the other case, the filter produces two non-zero samples (Figure~\ref{dec_resp_fig} (right)). 
The ratio of these samples depends on the phase difference between the input pulse and the ADC clock. 
Since this ratio is a monotonic function of the phase shift, it 
can be used to determine the arrival time of the input pulse. 
The amplitude of the input pulse is obtained from the sum of two non-zero samples multiplied by a 
time-depended correction-factor. 
All these operations can be done using look-up tables, possibly offline.

\section{Beam-test results}

In 2010, several million triggers have been recorded for two sensor
areas, on either a LumiCal or a BeamCal detector plane equipped with sensor
and FE  ASICs. In 2011, a similar number of triggers has been recorded
with complete detector planes equipped with sensor, FE ASIC,
ADC ASIC, and FPGA-based back-end.

\subsection{Study of a detector plane with a silicon sensor}

\begin{figure}[h!]
    \centering
  \subfigure[]{
      \includegraphics[width=0.450\textwidth]{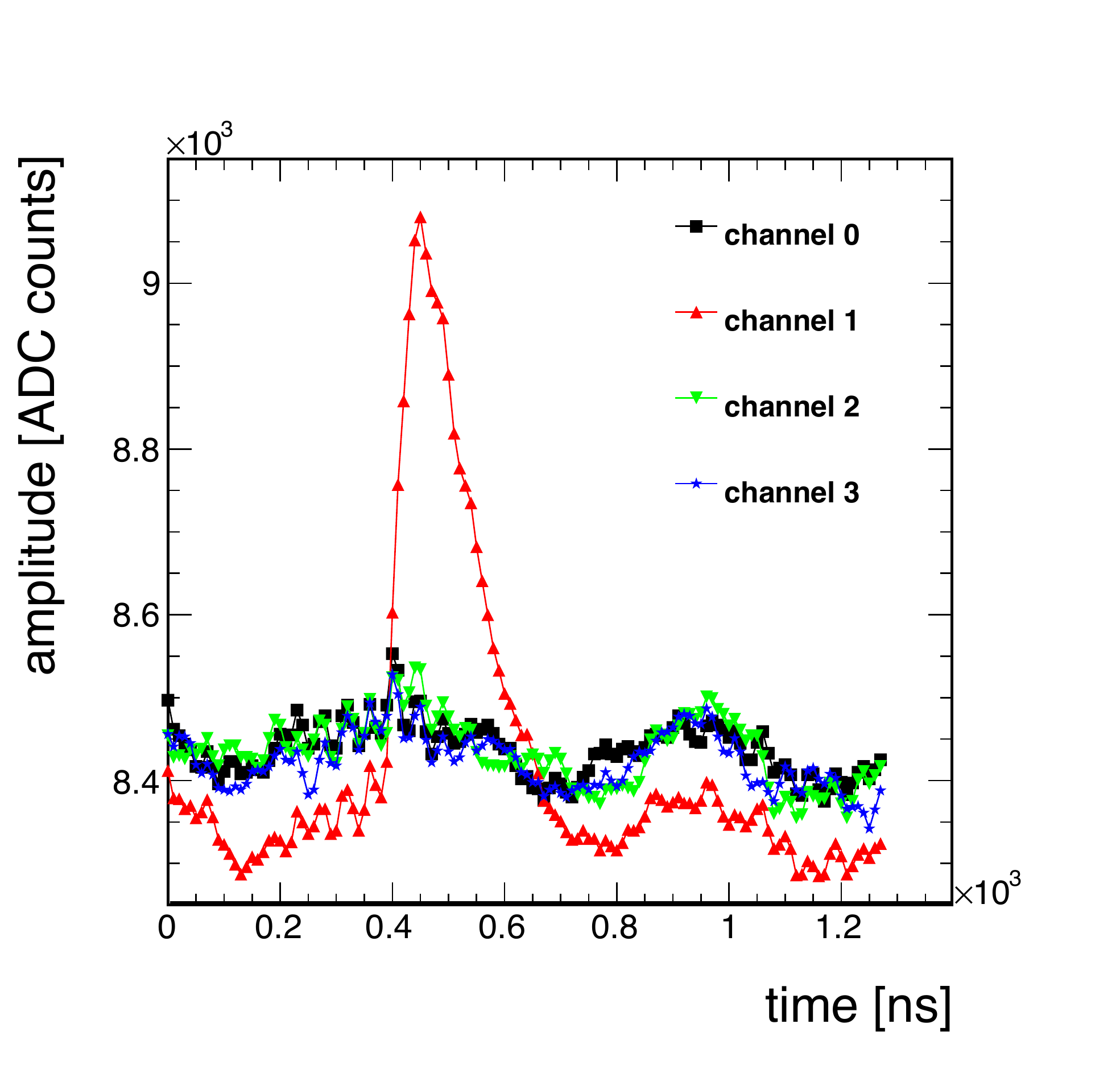}}
   \subfigure[] {
      \includegraphics[width=0.450\textwidth]{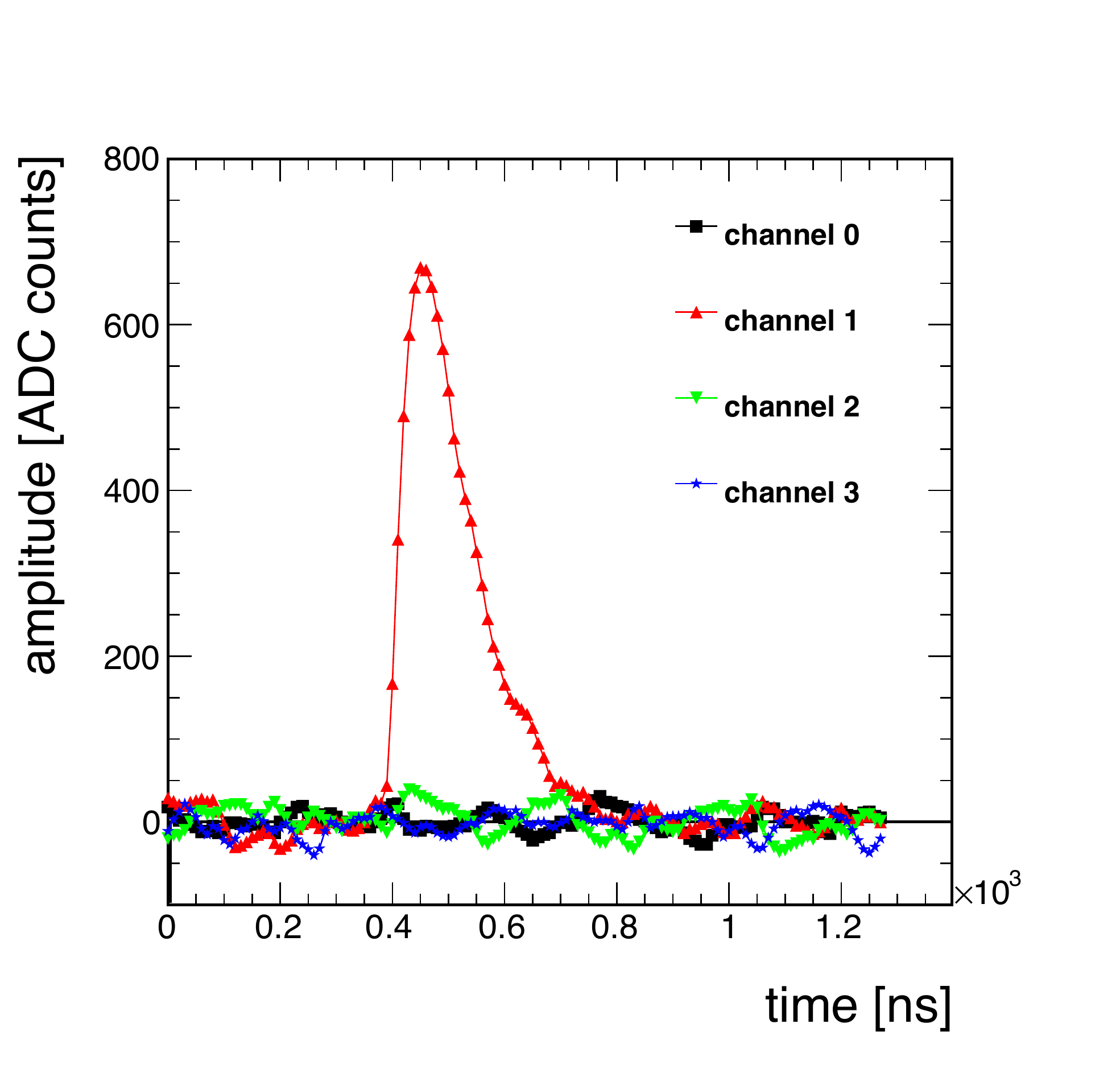}}
      \caption{ The digitised waveforms of 4 channels, (a) before  and (b) after common mode subtraction.\label{figure:both}}
 \end{figure}

During the 2010 beam test,  the analog signals were digitised by an external 14-bit ADC (CAEN V1724) with 100 MS/s.
The digitised waveforms of 4 channels readout in parallel are shown in Figure~\ref{figure:both}~\cite{Itamar}.

Common Mode Noise (CMN) is determined and subtracted from the raw data, taking into account the different front-end gain for the channels with $MOS$ and $R_f$ feedback~\cite{OlgaThesis,Szymon-thesis}.

\subsubsection{Amplitude spectrum and signal-to-noise ratio}

The amplitude is measured with respect to the baseline. Firstly, the average of 32 samples before the arrival of the pulse was used to estimate the baseline mean value,
as well as the baseline RMS for each channel, defining the noise. The baseline mean value was
subtracted from all samples, shifting the average baseline to zero. The amplitude is obtained as the maximum value of the signal. 

\begin{figure}[h!]
 \begin{center}
  \includegraphics[width=0.6\textwidth]{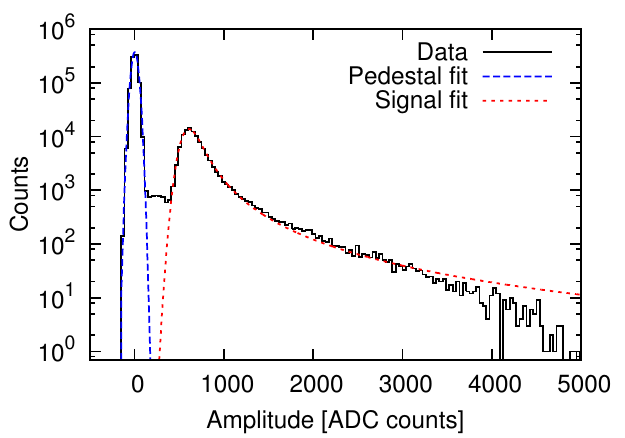}
  \caption{ Amplitude spectrum obtained from a silicon sensor pad used in the 2010 beam-test period after common mode subtraction
  and rejection of events with position close to pad borders. }
  \label{fig:tb3_spectra_raw}
 \end{center}
\end{figure}

The amplitude spectrum shown in  Figure~\ref{fig:tb3_spectra_raw} for the 2010 data,  after common mode subtraction and rejection of events with position close to pad borders, 
is fit well by a convolution of a
Landau distribution and a Gaussian for the signal and another Gaussian for the pedestal. The spectrum for the 2011 data is very similar.
The amplitudes in the pedestal peak are obtained when the
acquisition is triggered and no particle is crossing the considered pad. 

Using the most probable value (MPV) of the signal amplitude distribution, the signal-to-noise values are determined for all channels 
as
\begin{equation}
 S/N = \frac{\mathrm{MPV}}{\sigma(Pedestal)}.
\label{equ:SNR}
\end{equation}

The obtained $S/N$ values are in the range 16-22 when measured before the CMN subtraction, and 28-33 after it, for the prototype equipped with sensor and FE ASICs. For the complete detector prototype, comprising ADC ASICs, a $S/N$ in the range 11-15 was measured before, and 19-23 after CMN subtraction ~\cite{Szymon-thesis}. The detector prototype with ADC ASICs had significantly longer fanouts and a large connector between the sensor and readout boards.

\subsubsection{Crosstalk}

\begin{figure}[h!]
\centering
\includegraphics[width=1\textwidth]{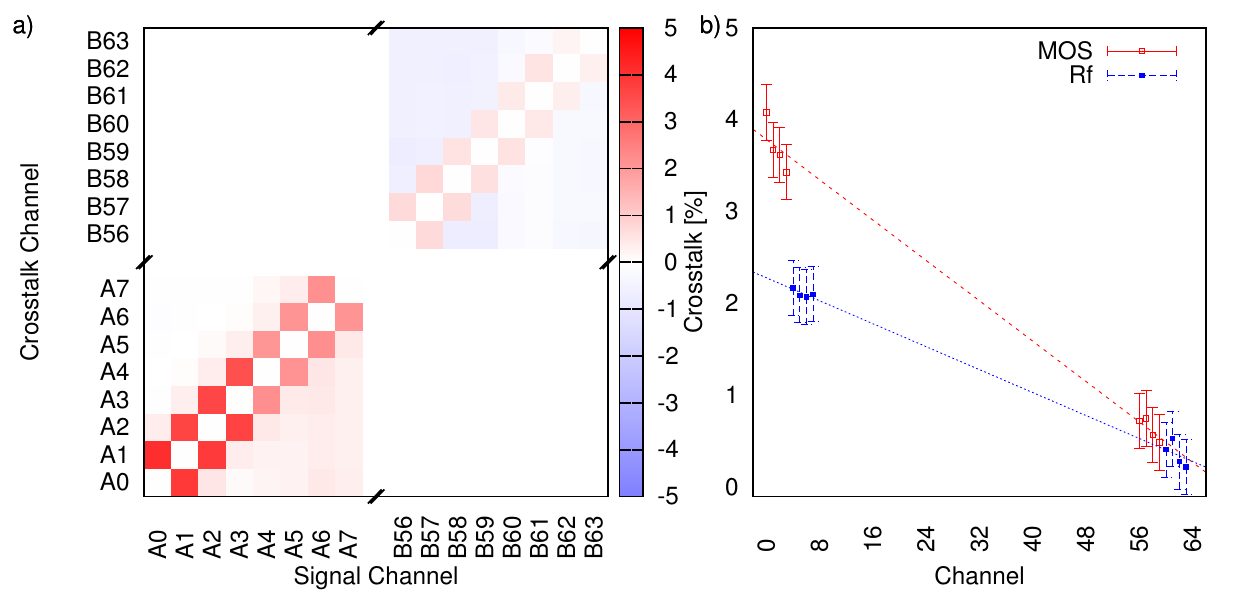}
\caption{a) Measured crosstalk coefficients for groups of eight pads. b) Measured crosstalk coefficients
to the closest neighbor as a function of pad number. The error bars represent statistical uncertainties only.}
\label{fig:cross_talk}
\end{figure}

Signal amplitudes on neighbouring signal pads were studied to evaluate the crosstalk level between pads. Only events with track impact points  identified to lie further than 200 $\mu$m from the pad border were used, to minimise the effect of charge sharing. 
The crosstalk coefficients between all channels are presented in Figure~\ref{fig:cross_talk}(a).
One can see a positive correlation between amplitudes of neighbouring channels, that can be interpreted as crosstalk.
For the crosstalk measurement, the CMN subtraction procedure was not applied to avoid artificially introduced biases. 
Significant crosstalk can be observed only between the closest neighbors. In Figure~\ref{fig:cross_talk}(b) the estimated crosstalk for pads with long fanout traces (1 - 8) and short fanout traces (56 - 64) is shown.
The highest crosstalk is observed for pads with the longest fanout traces (A0 - A7) attached to it. Moreover, there is a
significant difference between channels with different feedback type, which can be explained by
the differences in the input impedance. Since the effect and the difference between feedback types is not negligible for the innermost channels, its impact on the spatial and energy resolution for showers needs to be  quantitatively simulated in the future.

\subsubsection{Signal deconvolution}

\begin{figure}[!h]
    \centering
    \includegraphics[width=1.\textwidth]{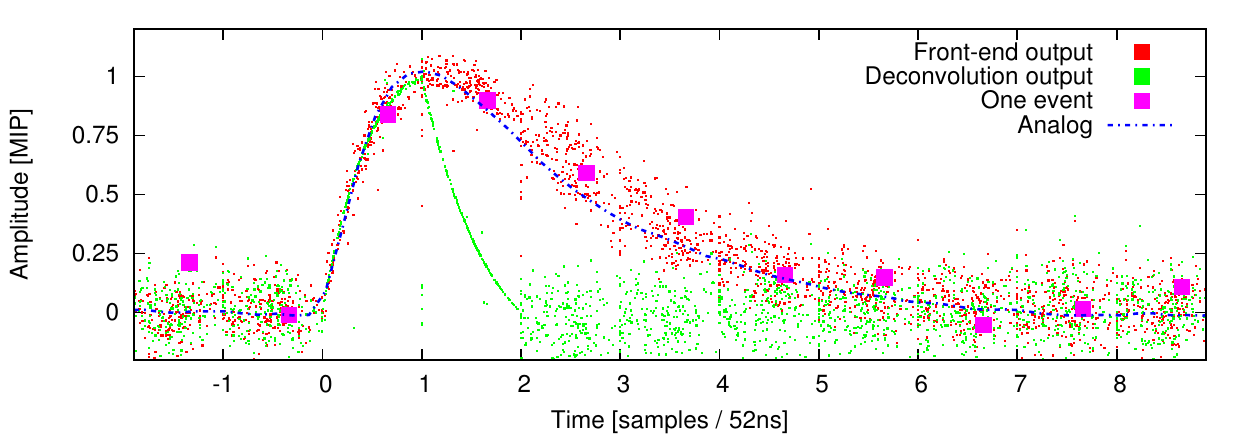}
    \caption{The amplitude as a function of time before (red) and after deconvolution (green).
    The pulses were time-aligned using the time information obtained from the
    deconvolution algorithm. For illustration, an example event is shown as squares.}
    \label{fig:tb2_event_waveforms}
\end{figure}

In 2011, the data were collected in two modes of operation, synchronous and asynchronous with the beam clock. An example of measured points collected in the asynchronous mode is
presented in Figure~\ref{fig:tb2_event_waveforms}.
To extract information about the pulse amplitude and time, the deconvolution method,
as described in Section~\ref{chp:deconvolution}, was applied. 
The waveform, representing an
analog output of the FE electronics captured with a very fast digital
oscilloscope, is shown as a reference. The events were precisely time-aligned
using the information about the time obtained from the deconvolution processing
algorithm. The pulse duration after deconvolution is as expected shorter  and the amplitude is correctly reconstructed. The $S/N$ ratio after deconvolution was similar to the one obtained without deconvolution, running the system in synchronous way.

\subsubsection{Response as a function of position in the sensor}

\begin{figure}[h!]
\centering
\includegraphics[width=0.37\textwidth]{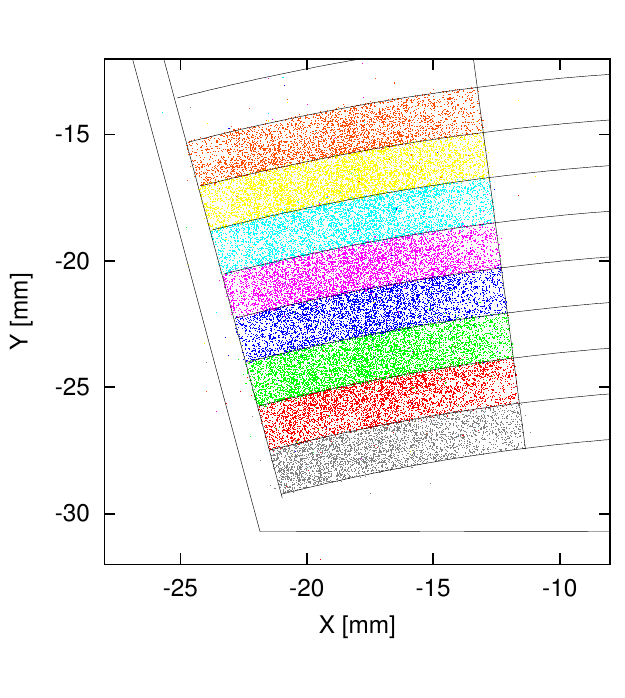}
\caption{Reconstructed position of impact points on the sensor with color assignment determined by the pad with signals above threshold, registered in the LumiCal detector plane.}
\label{fig:position}
\end{figure}
Using a track fit from the MVD telescope, the impact position of the beam electron on the sensor is calculated. The distribution of hits is shown in Figure~\ref{fig:position}. 
The colors, characterising the pads, are assigned to each point in case the signal in the
pad is above a certain threshold. The pad structure for the LumiCal sensor is  reproduced correctly.

\begin{figure}[h!]
\centering
\includegraphics[width=0.78\textwidth]{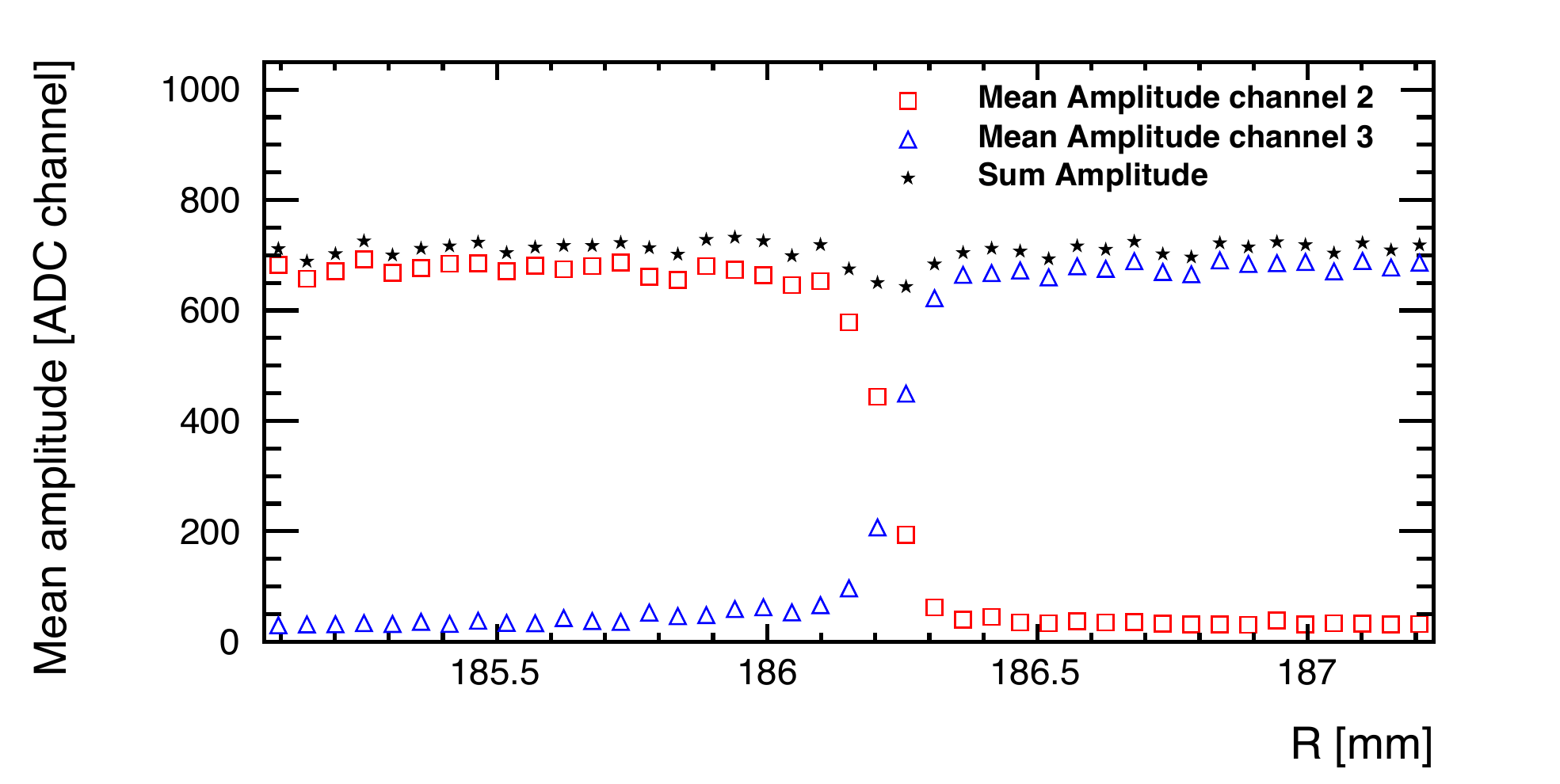}
\caption{The mean of the signal amplitude in a single pad as a function of the hit position on the border between two adjacent pads (channels 2 and 3). The sum of the signals in the two neighbouring pads is shown as stars. The statistical uncertainty is less or equal to the size of the symbols.
}
\label{fig:pad_gap}
\end{figure}

In Figure~\ref{fig:pad_gap} the mean amplitude is shown as a function of the position across two adjacent pads. A drop of about 10\% is observed in the 100 $\mu$m gap between pads. 

\subsubsection{Shower development}

The electron shower development as a function of the  absorber depth in front of the sensor plane was measured. Tungsten absorber plates of 3.5~mm thickness, corresponding roughly to one radiation length each, were positioned in front of the sensor plane.
For each configuration, 50\ 000 events were collected and analysed. Given the availability of only one sensor and the  stochastic character of the  electromagnetic shower development, only average quantities were analysed.  For these studies, the electron beam was directed to the centre of the
instrumented area of the sensor plane. For each absorber thickness, Monte-Carlo (MC) expectations using the GEANT4 framework were carried out. 
The electron beam was modeled as isotropically distributed with trajectories
collimated into a square shape corresponding to the beam size.

\begin{figure}[h!]
  \centering
  \includegraphics[width=1.0\textwidth]{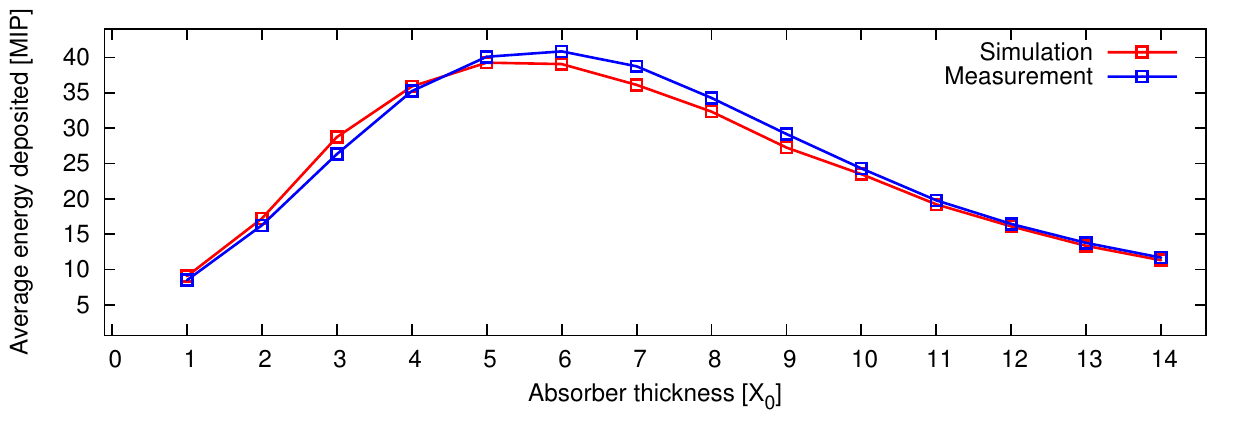}
  \caption{Average energy deposited in the instrumented area of the LumiCal sensor as a function of
  the tungsten absorber thickness, expressed in radiation length, ${\rm X_0}$.The statistical uncertainty is less or equal to the size of the symbols.}	
  \label{fig:tb2_shower_mean} 
\end{figure}

The measured average energy deposited in the instrumented area as a function of the
tungsten plate thickness is shown in Figure~\ref{fig:tb2_shower_mean}~\cite{Szymon-thesis}. The  shower data were recorded at an electron energy of 4.6~GeV.
The result of the MC simulation is in reasonable agreement with the measurement. The shower maximum is observed after six radiation lengths.

\subsection{Study of a detector plane with a GaAs sensor}

\begin{figure}[h!]
  \centering
   \includegraphics[width=0.5\columnwidth]{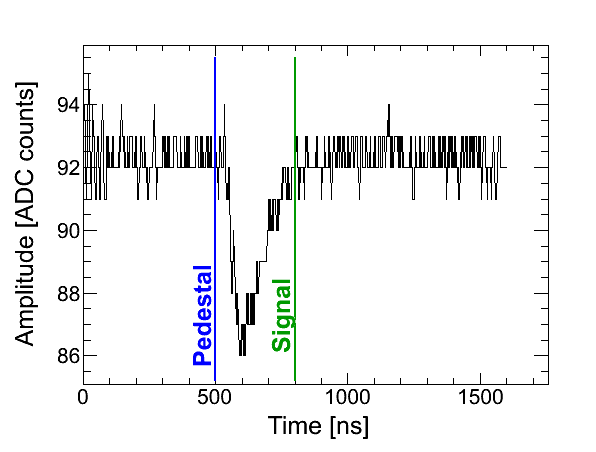}
   \caption{A digitised signal of one channel of the GaAs detector plane recorded with a 500 MS/s ADC.}
    \label{fig:B_tb1_signal.jpg}
\end{figure}

\begin{figure}[h!]
 \begin{minipage}[t]{0.5\linewidth}
  \centering
    \includegraphics[width=1\columnwidth]{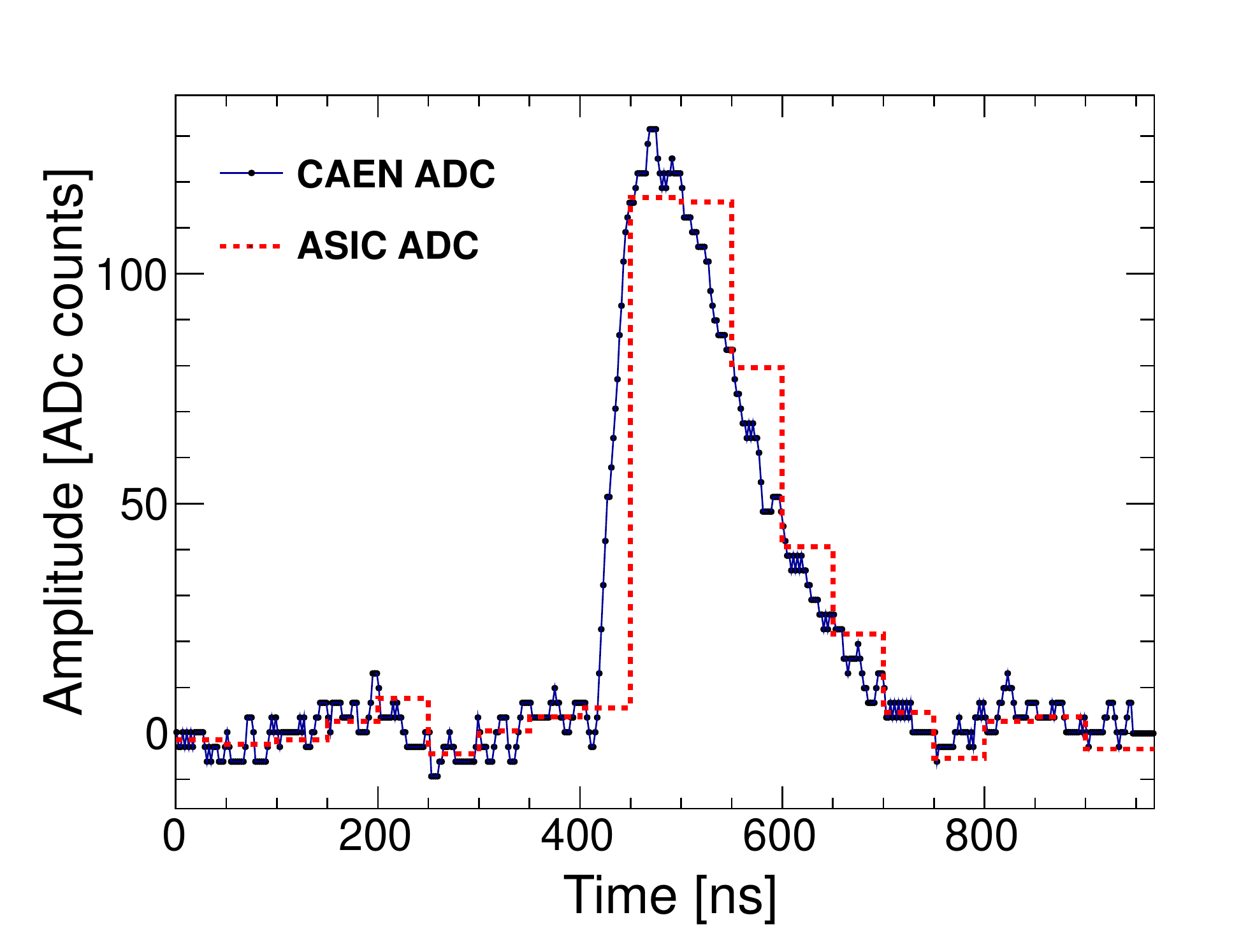}
    \caption{An example of the signal digitised by the external 500~MS/s CAEN ADC (full line) and the 20~Ms/s ADC ASIC (dotted line).}
    \label{fig:TB11Signal}
  \end{minipage}
\hspace*{0.07\linewidth}
  \begin{minipage}[t]{0.5\linewidth}
   \centering
    \includegraphics[width=1\columnwidth]{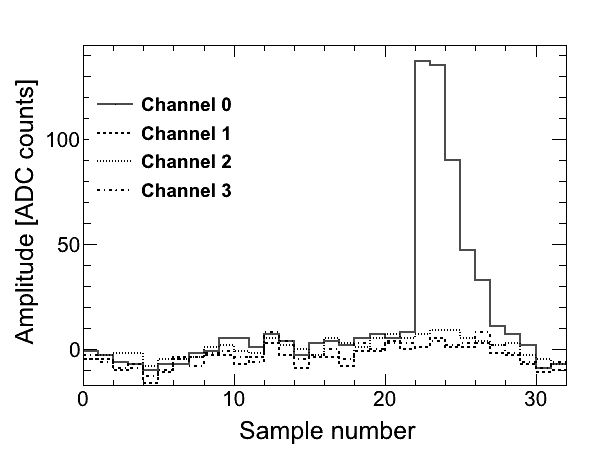} 
    \caption{Raw signal shapes of four adjacent channels digitised with the ADC ASIC.}
    \label{fig:raw_signal_all_samples}
  \end{minipage}
\end{figure}

In 2010, the signals from the FE ASICs were digitised by an external 
8-bit flash ADC V1721 performing at 500 MS/s.
An example of a signal is shown in Figure~\ref{fig:B_tb1_signal.jpg}. 
For each trigger the signal was digitised in a time window of 1.6 $\mu$s. In the first 500~ns, the average baseline and its RMS were evaluated.
The adjacent time window of 300 ns is used to analyse the signal.  
During the 2011 beam test, the signal was split after the FE ASIC
and simultaneously digitised by the external ADC (8-bit) and the ADC ASIC (10-bit) on 
the readout board. Both results are shown in Figure~\ref{fig:TB11Signal}, and found to exhibit the same 
shape.
The raw signal shapes of four adjacent channels of the ADC ASIC are shown in Figure~\ref{fig:raw_signal_all_samples}. 
Small CMN is observed which is subtracted from all raw data, taking into 
account different gains in the channels.  

\subsubsection{Amplitude spectrum and signal-to-noise ratio} 

\begin{figure}[h!]
   \centering
    \includegraphics[width=0.45\columnwidth]{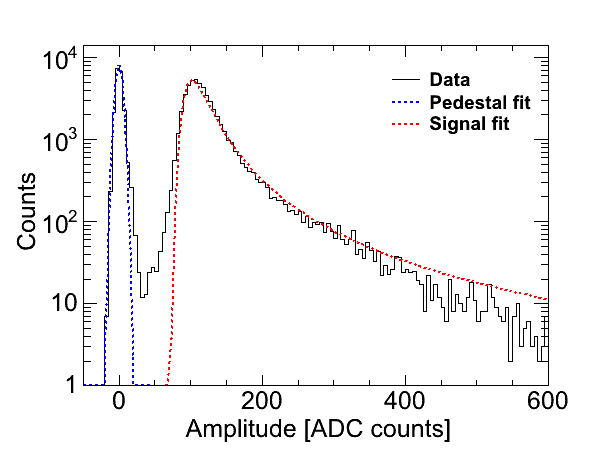}
     \caption{The amplitude spectrum of the GaAs detector plane at a bias voltage of 100~V.}
      \label{fig:B_tb1_spectrum.jpg}
\end{figure}

\begin{figure}[h!]
 \begin{minipage}[t]{0.5\linewidth}
  \centering
    \includegraphics[width=0.9\columnwidth]{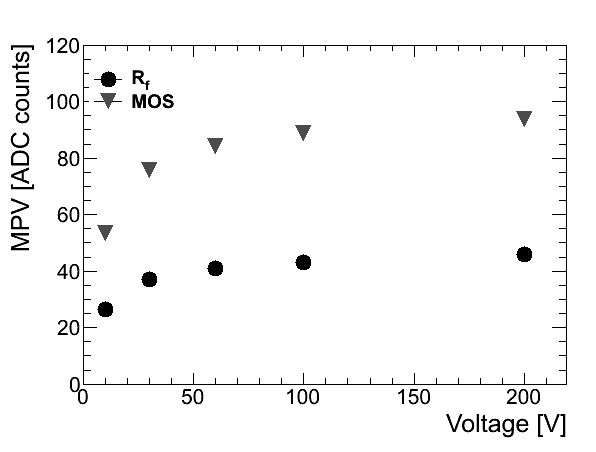}
    \caption{The signal amplitude as a function of the applied bias voltage measured using FE ASICs with $R_f$ and $MOS$ feedback. The statistical uncertainty is less or equal to the size of the symbols. From the calibration coefficient a fully correlated systematic uncertainty of 5\% is estimated.}
    \label{fig:TB11_MIP_HV}
   \end{minipage}
\hspace*{0.02\linewidth}
  \begin{minipage}[t]{0.5\linewidth}
   \centering
    \includegraphics[width=0.9\columnwidth]{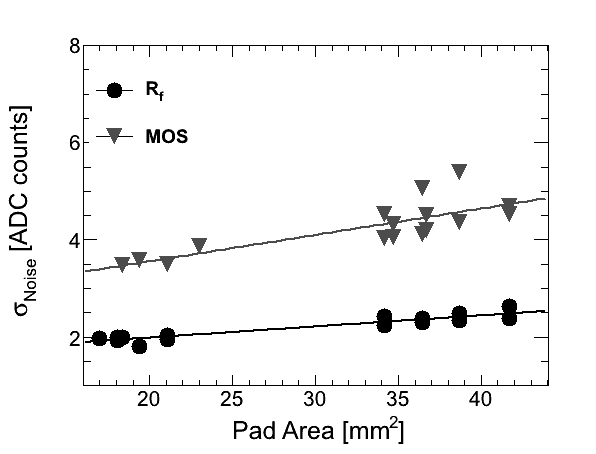}
    \caption{The pedestal standard deviation, $\sigma_{Noise}$,  as a function of the pad area for $R_f$ and $MOS$ feedback. The statistical uncertainty is less or equal to the size of the symbols.}
    \label{fig:TB11_NOISE_PAD}
   \end{minipage}
\end{figure}

The signal amplitude is defined as the difference between the maximum ADC value in the signal time window, 
as shown in Figure~\ref{fig:B_tb1_signal.jpg}, and the average baseline.
As an example, an amplitude spectrum is shown in Figure~\ref{fig:B_tb1_spectrum.jpg} for a bias voltage of 100~V.
The distributions of the amplitude at each value of the bias 
voltage are fitted with a convolution of a Gaussian and a Landau
distribution to determine the MPV.
Figure~\ref{fig:TB11_MIP_HV} shows the 
MPV values 
as a function of the bias voltage for both feed-back types used in the FE ASICs. 
The MPV grows slowly with increasing voltage and reaches saturation
above 60 V, in agreement with the laboratory measurements shown in Figure~\ref{figure:gaas_cce_vs_hv_hep}.
Taking into account the energy loss of beam electrons in the sensor, and the  calibration of the readout chain, the CCE approaches
42.2~\% at 100 V bias voltage, in reasonable agreement
with laboratory measurements~\cite{OlgaThesis} when taking into account a 5\% uncertainty of the calibration factors.

The $S/N$ ratio was measured using the 
MPV values and the standard deviation of the pedestal distributions. As an example, Figure~\ref{fig:TB11_NOISE_PAD} shows the standard deviation of the pedestal  
using the amplitude from the 
32 readout pads as a function of the pad size. As expected, 
the noise depends almost
linearly on the pad size for both FE feedback types. It is higher for the FE ASICs with $MOS$ feedback than that with $R_{\it{f}}$ feedback due to the higher gain of the $MOS$ feedback in the calibration mode.

\begin{figure}[!ht]
   \centering
    \includegraphics[width=0.55\columnwidth]{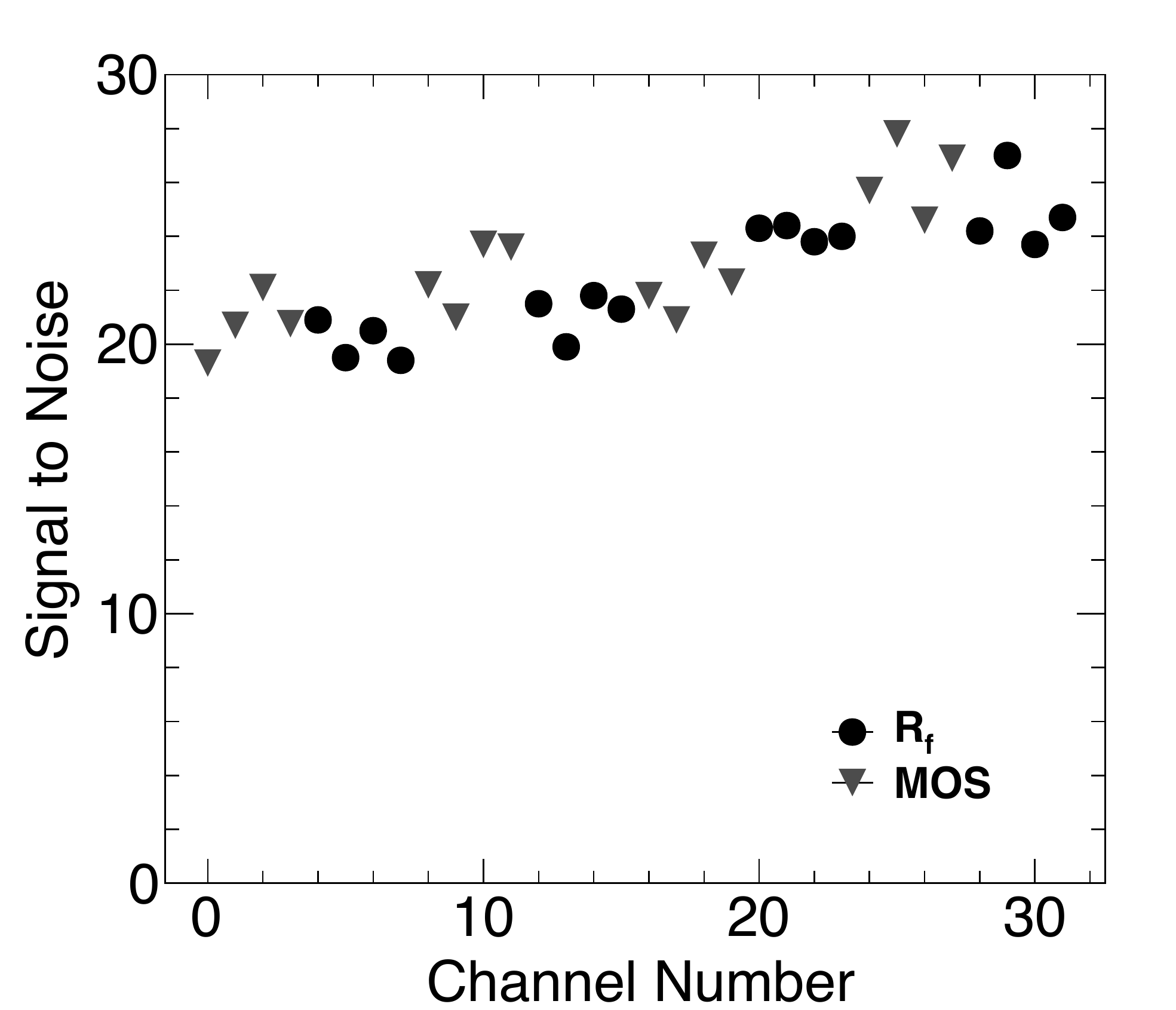}
    \caption{The $S/N$ ratios, as a function of the channel number for both  $R_{\it{f}}$ and $MOS$ feedback. The statistical uncertainty is less or equal to the size of the symbols.}
    \label{fig:S/N_ampl_int}
\end{figure}

The $S/N$ values, determined using Eq. (\ref{equ:SNR}) for all channels, 
are shown in Figure~\ref{fig:S/N_ampl_int}.  Though the signal amplitudes and the noise are different for the two feedback schemes, their ratio is very similar. 
The slight slope is due to the larger pad areas at low channel numbers. The variation within a chip reflects the range of response spread.

\subsubsection{Signal deconvolution} 
\label{chp:beamcal_deconvolution}

\begin{figure}[h!]
  \centering
  \subfigure[]{
  \includegraphics[width=0.48\linewidth]{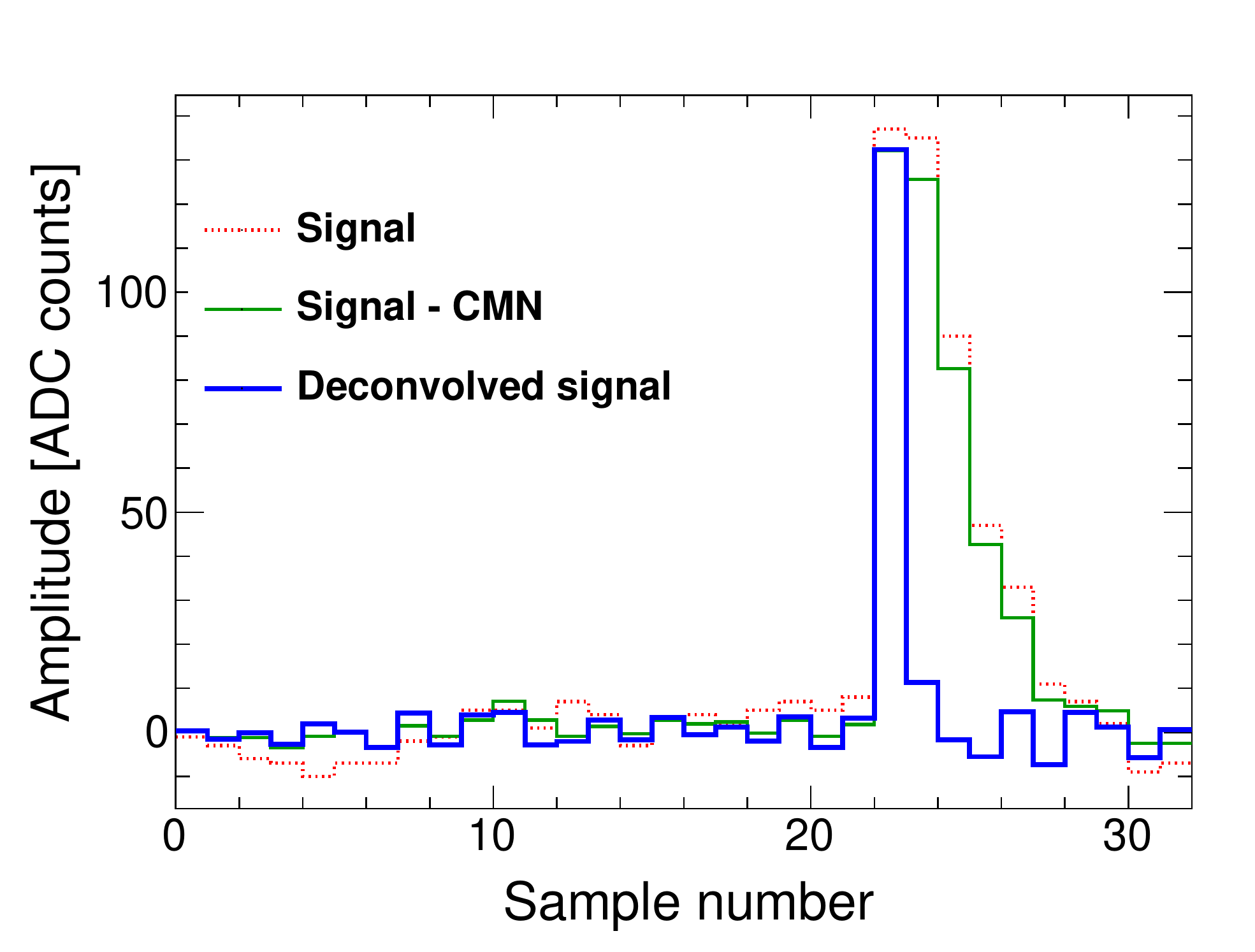}
  \label{fig:TB11_deconv_example}}
  \subfigure[] {
  \includegraphics[width=0.48\linewidth]{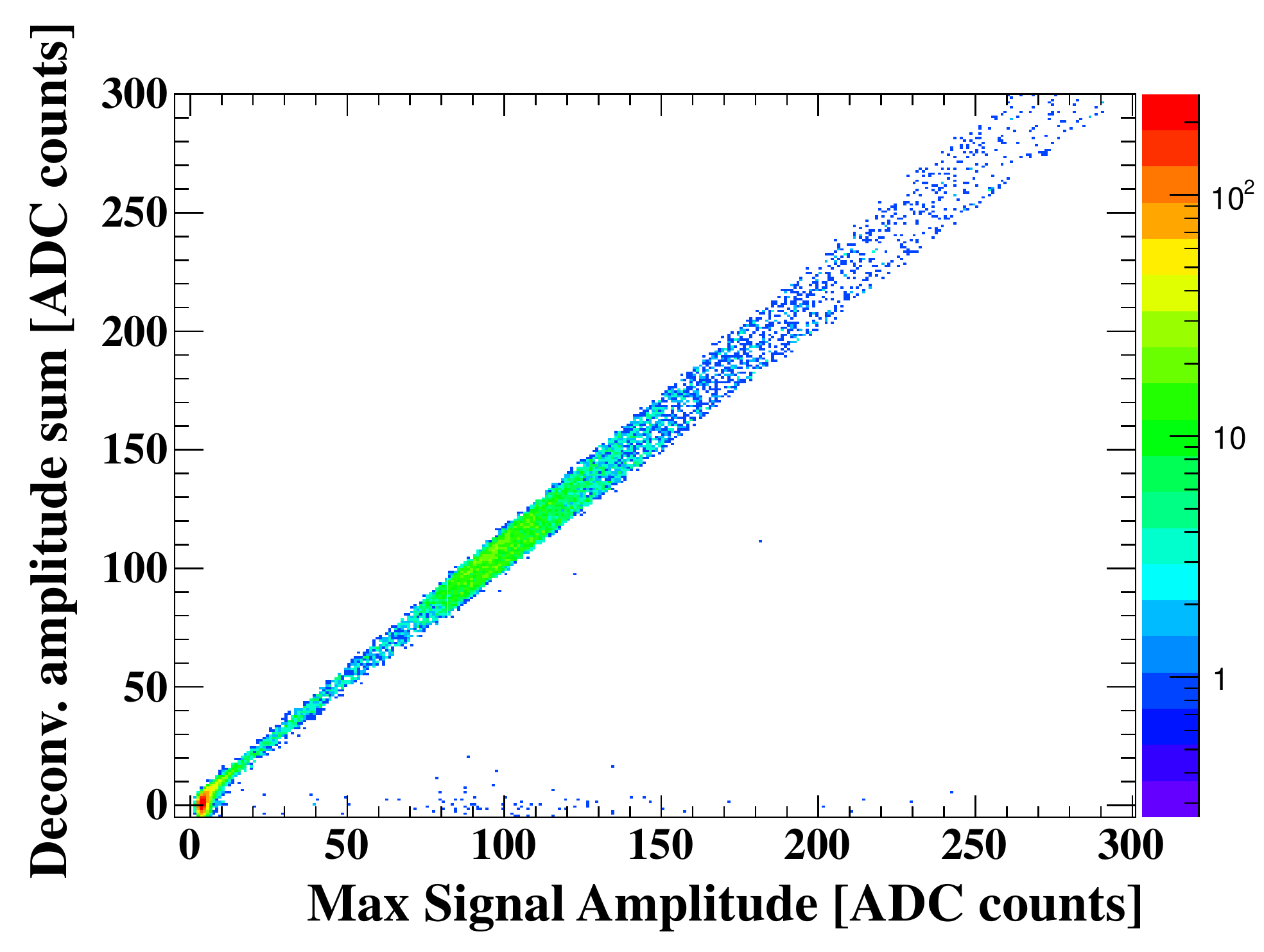}
  \label{fig:TB11CorrAmpl22Deconv22}}
  \caption{(a) The amplitude in channel 0 as a function of time digitised by the ADC ASICs (red), after the CMN subtraction (green), after deconvoluted (blue).
(b) The sum of the two first non zero amplitudes after the deconvolution filter, as a function of the signal amplitude.}
\end{figure}

Similarly to the LumiCal test (see Section~\ref{chp:deconvolution}), data were collected in synchronous and asynchronous modes and deconvolution was performed. 
Figure~\ref{fig:TB11_deconv_example} shows the raw signal after CMN subtraction and the deconvoluted signal.
For the time constant and the sampling interval of the ADC, $\tau~=$~60~ns  and $T_{smp}$=50~ns were used, respectively.
The pulse after deconvolution is significantly shorter.

Figure~\ref{fig:TB11CorrAmpl22Deconv22} shows the correlation between the amplitude measured with the ADC and the amplitude obtained after deconvolution in the asynchronous mode. A linear dependence is observed. For the synchronous operation mode, the deconvoluted amplitudes show the same linear dependence with a slightly smaller spread~\cite{OlgaThesis}.

\subsubsection{Response as a function of position in the sensor}

\begin{figure}[h!]
\centering
\includegraphics[width=0.6\columnwidth]{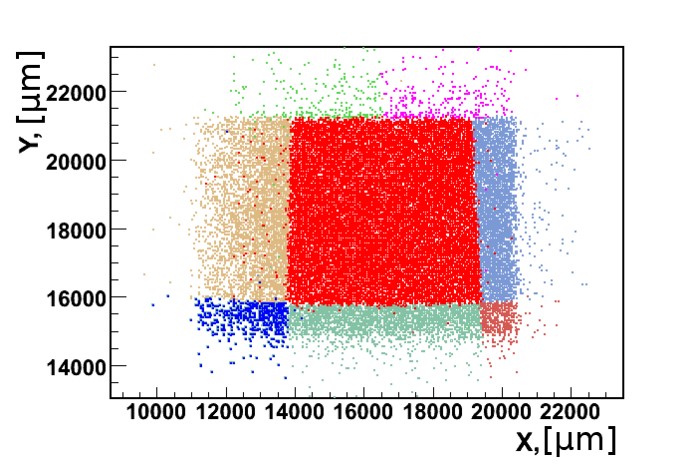}
\caption{Distribution of the impact points of beam electrons on the GaAs detector plane. A given 
color is assigned to each pad if the signal in that pad is above a certain threshold.   
}
\label{fig:beamcal_hitmap}
\end{figure}

\begin{figure}[h]
  \centering
  \includegraphics[width=0.9\columnwidth]{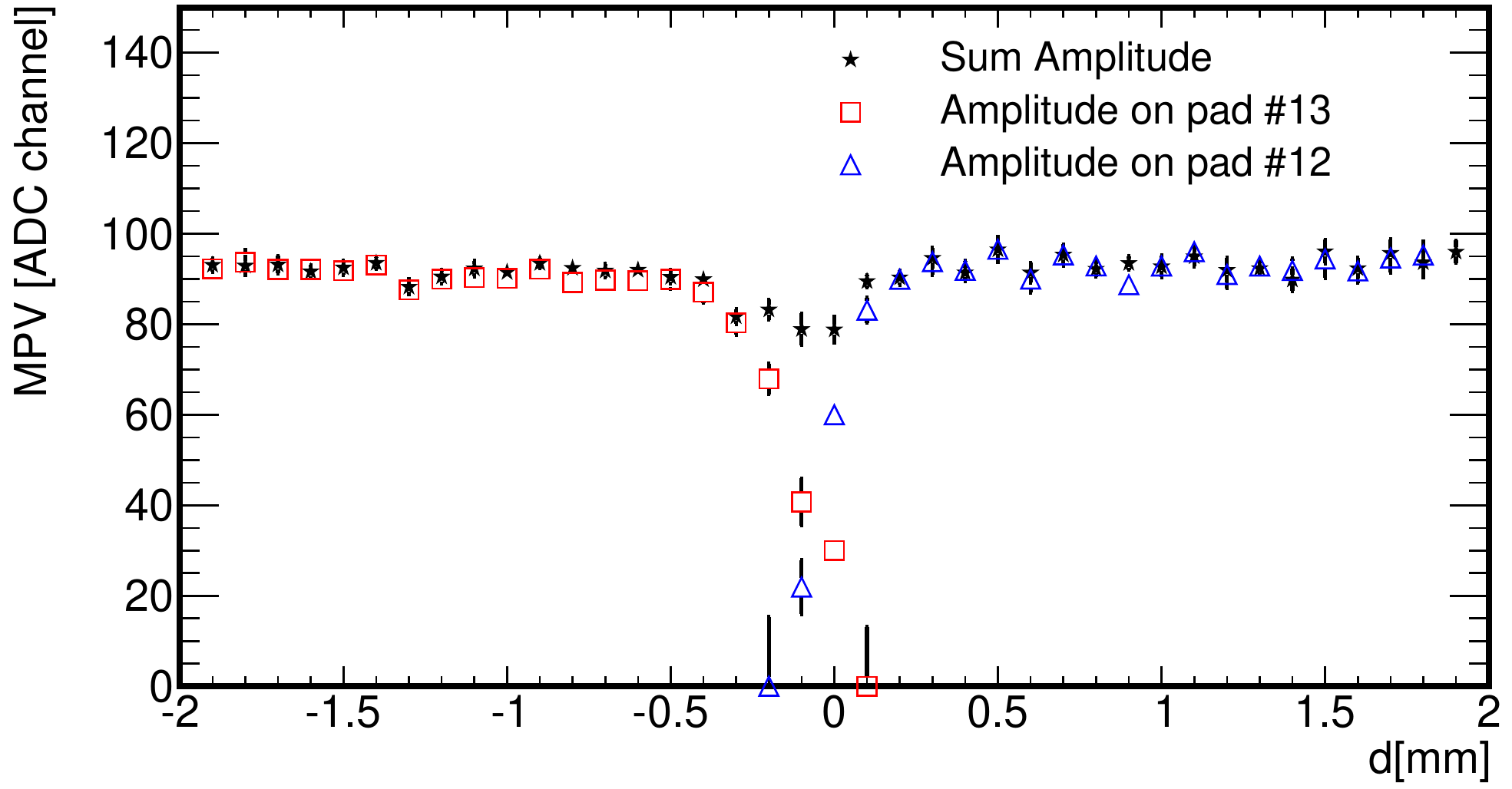} 
   \caption{ The MPV of the signal amplitude as a function of the hit position on the border between two adjacent pads. The sum of the signals in the two neighboring pads is shown as stars. The error bars represent statistical uncertainties.}
 \label{fig:MPV_vs_distance}
\end{figure}

Using the MVD telescope, the impact point of each beam electron on the detector plane is 
calculated. In case the signal of a pad is above a certain threshold, a pad-depending 
color is assigned to the impact point.
An example distribution of impact points is shown in Figure~\ref{fig:beamcal_hitmap}.
The pad structure of the GaAs BeamCal sensor is accurately reproduced in the plot.
From amplitude spectra for different areas on the pad, the MPV values are determined and found to be in
agreement within their statistical uncertainties.
The MPVs are also measured in slices of 100 $\mu$m covering the non-metallised gap between two pads.
For beam electrons
with impact points near and in the gap, sharing of the signal between the
adjacent pads is observed, as can be see in Figure~\ref{fig:MPV_vs_distance}. 
When the amplitudes on the two adjacent pads are added, their MPV as a function of the position exhibits
a drop of about 10\% in the region of the gap.

\section{Conclusions}

Detector plane prototypes for LumiCal and BeamCal were prepared and tested in an electron
beam with energies between 2 and 4.5 GeV. The detector planes comprised silicon or GaAs pad sensors, FE ASICs, ADC ASICs and an FPGA for orchestrating the readout. For both detectors, signal-to-noise ratios between 20 and 30 are obtained. The performance of FE ASICs with $R_f$ and $MOS$ feedback is very similar except for the cross talk. In certain channels the cross talk for the $MOS$ feedback is by a factor two larger than that of the $R_f$ feedback. A deconvolution method is applied to reduce the amount of recorded data, while keeping the timing and amplitude information of the signals. From studies of the detector response as a function of the beam-electron impact point, a uniform performance was found, apart from a 10\% reduction of the amplitude in the vicinity of the gaps between adjacent pads. A Monte-Carlo simulation of the shower shape as a function of the absorber depth shows reasonable agreement with the measurements. The measured sensor and readout electronics parameters 
allow for an efficient detection of minimum ionising particles in the calorimeters, necessary for sensor-plane alignment and frequent calibration of the response of all pads to account for performance degradation due to radiation damage. 

The results demonstrate the full functionality of the sensor planes both with silicon and GaAs sensors. The next step will be the construction of a calorimeter prototype comprising a compact multilayer structure of alternative sensor and absorber planes. It will have the possibility to operate using power pulsing.
Beam-test data taken with this prototype will be used to verify the performance of the reconstruction of electromagnetic showers estimated from Monte Carlo simulations.

\section*{Acknowledgments}

This work was supported by the Commission of the European Communities under the 6th and 7th Framework Programs EUDET and AIDA, contract no. 262025. The Tel Aviv University is supported by the German-Israel Foundation (GIF), the Israel Academy of Sciences and the I-CORE Program of the Planning and Budget Committee and the Israel Science Foundation (ISF) (grant No. 1937/12).  The Pointificia Universidad Catolica de Chile is supported by the Chilean Commission for Scientific and Technological Research (CONICYT) grant FONDECYT 11110165. The AGH-UST is supported by the Polish Ministry of Science and Higher Education under contract no. 2156/7.PR UE/2011/2. The  IFIN-HH was supported by the Romanian UEFISCDI agency under PCE-ID\_806 and by the Ministry of Education and Research under PN 09 37 01 01. The INP PAN was supported by the Polish Ministry of Science and Higher Education under contracts 141/6.PR UE/2007/7, 2369/7.PR/2012/2 and by the EU Marie Curie ITN, grant number 214560. The ISS was supported by the Romanian UEFISCDI agency under PN-II-RU-TE-2011-3-0279 and by the Romanian Space Agency (ROSA), contracts STAR-C1-1 44/2012, STAR-C2 64/2013 and STAR-C2 85/2013. The Vinca Institute of Nuclear Sciences has been partially funded by the Ministry of Science and Education of the Republic of Serbia under the project Nr. OI 171012.

\end{document}